\documentclass[%
reprint,
superscriptaddress,
 amsmath,amssymb,
 aps,
prb,
]{revtex4-2}

\usepackage{graphicx}
\usepackage{dcolumn}
\usepackage{bm}
\usepackage{hyperref}

\usepackage{xcolor}

\newcommand{\G}{\mathcal{G}}

\begin{document}

\preprint{APS/123-QED}

\title{Electrical probe of spin-spiral order in quantum spin Hall/spin-spiral magnet van der Waals heterostructures}

\author{Fedor Nigmatulin}
\email{fedor.nigmatulin@aalto.fi}
\affiliation{QTF Centre of Excellence, Department of Electronics and Nanoengineering, Aalto University, FI-00076 Aalto, Finland}

\author{Jose L. Lado}
\affiliation{Department of Applied Physics, Aalto University, FI-00076 Aalto, Finland}

\author{Zhipei Sun}
\affiliation{QTF Centre of Excellence, Department of Electronics and Nanoengineering, Aalto University, FI-00076 Aalto, Finland}

\date{\today}

\begin{abstract}
Two-dimensional spin-spiral magnets provide promising building blocks for van der Waals heterostructures due to their tunable spin textures and potential for novel functionalities for quantum devices. However, due to its vanishing magnetization and two-dimensional nature, it is challenging to detect the existence of its noncollinear magnetization. Here, we show that a van der Waals junction based on a spin-spiral magnet and a quantum spin Hall insulator enables obtaining signatures of noncollinear magnetization directly from electrical measurements. Our strategy exploits the sensitivity of helical states to local breaking of time-reversal symmetry, potentially enabling the detection of local magnetic orders even in the absence of net magnetization. We show that the combination of spin-spiral order and nonmagnetic disorder gives rise to scattering in the helical channels that can be directly associated with the spiral exchange coupling and residual nonmagnetic disorder strength. Our results show how electrical transport measurement may offer a way to detect spin-spiral magnets by leveraging helical states in van der Waals heterostructures.
\end{abstract}

\maketitle

\section{Introduction}
Frustrated and noncollinear magnetic systems provide an exciting platform for exploring fundamental phenomena and the great potential for stray-field-free spintronics \cite{bousquet2016, steinbrecher2018, Bergmann2014, Menzel2012, rimmler2024, Nigmatulin2021}. 
Spin-spiral magnets (SSMs) feature a variety of exotic phenomena, including multiferroicity \cite{Tokura2010, bousquet2016, Song2022}, unconventional spin excitations \cite{tokura2021, Wuhrer2023}, and topological magnetism \cite{Fert2017}. The detection of noncollinear magnetic states in three-dimensional bulk systems can be achieved with neutron scattering \cite{furrer2009, muhlbauer2019}. However, probing spin spirals in two-dimensional (2D) materials represents a bigger challenge due to their reduced cross section. 
Spin-polarized scanning tunneling microscopy \cite{Bergmann2014, bode2007} 
requires that the magnet is the topmost layer and thus becomes challenging in encapsulated samples. Detecting symmetry breaking with vanishing magnetization can enable characterizing magnetic 2D materials, including monolayer spin-spiral multiferroic NiI$_2$ \cite{Song2022, Fumega2022, PhysRevLett.131.036701, PhysRevB.106.035156,Amini2024}, van der Waals heavy-fermion Kondo lattices \cite{Posey2024, Fumega2024, PhysRevB.109.L201118, Vano2021, Wan2023}, and
frustrated spin liquid candidates like 1T-TaS$_2$ \cite{MaasValero2021, Wang2024, Law2017, Chen2022, Ruan2021}. The conductance of the topological helical edge states coupled with a 2D magnetic material may offer an appealing strategy to tackle this problem.

\begin{figure}[t!]
    \centering
    \includegraphics[width = 8.6cm]{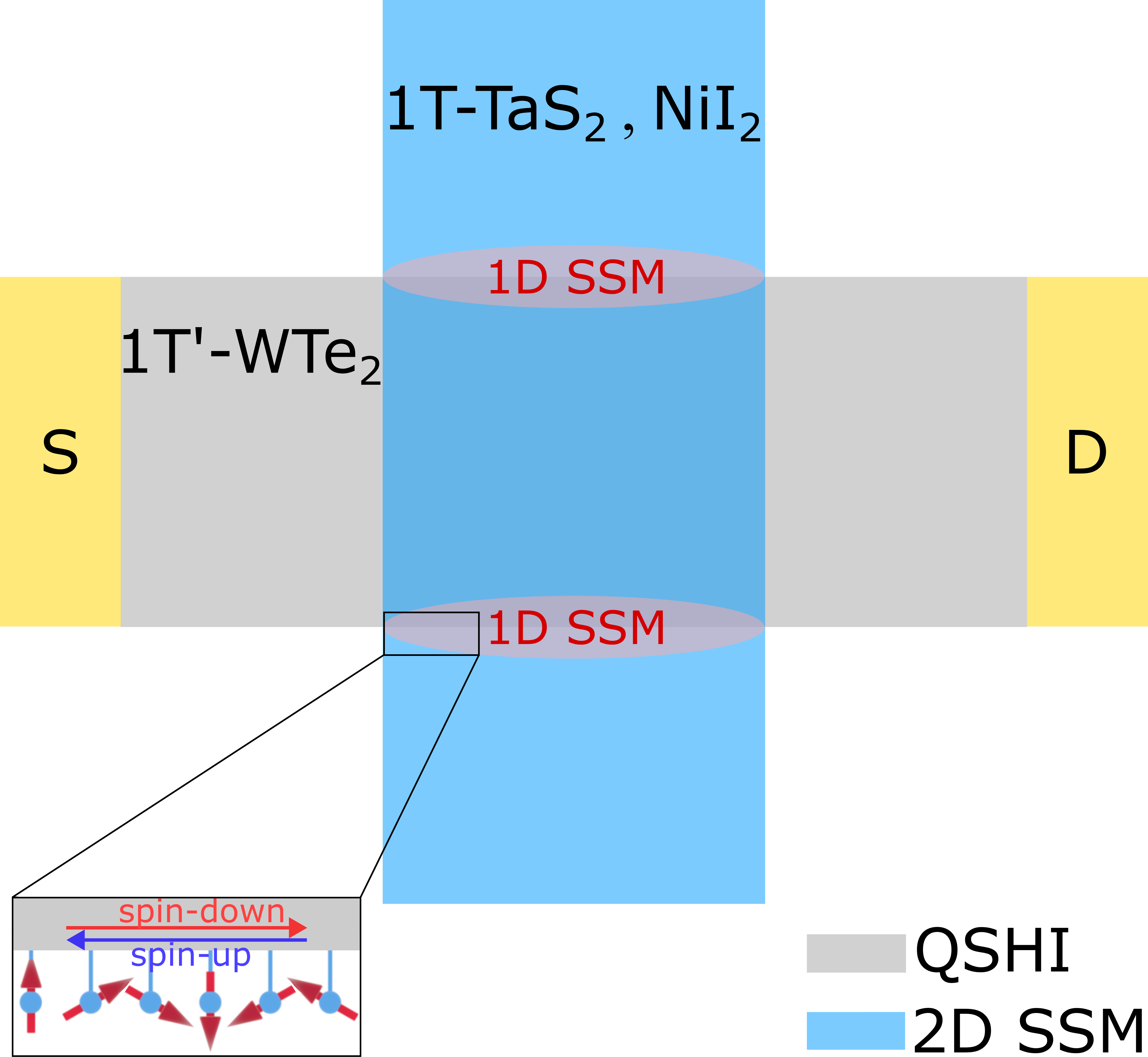} 
    \caption{Schematic of the van der Waals heterostructure formed by a QSHI (1T$^{\prime}$-WTe$_{2}$ monolayer) and a 2D SSM (1T-TaS$_{2}$ or NiI$_2$ monolayer). Exchange proximity leads to 1D spin spirals on the edges of 1T$^{\prime}$-WTe$_{2}$ flake. “S” and “D” denote source and drain electrodes, respectively, providing a current through the channel. Red arrows in the inset show local magnetic moments with angle $\pi/3$ between neighboring sites ($q = 1/6$), while the thin horizontal blue and red arrows represent spin-polarized edge states.}
    \label{fig:fig1}
\end{figure}

Topological insulators (TIs) \cite{BHZ2006, konig2007, xia2009, fei2017} are promising constituents of future technologies such as topological quantum computing \cite{fu2008} and spintronics \cite{he2022}.
 The potential of TIs stems from the existence of gapless edge states protected as long as a specific symmetry is not broken. Among topological states, quantum spin Hall insulators (QSHIs) such as 1T$^\prime$-WTe$_2$ \cite{Tang2017,Wu2018} provide a potential platform for dissipationless spin transport and fundamental building blocks to engineer other exotic states by creating artificial interfaces \cite{Ueda2013, Yokoyama2010, li2013}, including Chern insulators and topological superconductors \cite{fu2008, Teixeira2019, Li2015}.
In the case of quantum spin Hall insulators, the topological edge modes feature spin-momentum locking and robustness to nonmagnetic external perturbations that do not break time-reversal symmetry (TRS) protection. 

Here, we show that the electronic transport of helical QSHI states could potentially be employed as a probe of noncollinear magnetism in QSHI/SSM van der Waals heterostructures. We demonstrate that the simultaneous presence of spin magnetization and nonmagnetic disorder quenches the conductance of topological edge states. While some effects of helical magnetic structures on topological surface states, such as modification of the band structure and group velocity of electrons, have already been studied previously \cite{Stagraczynski2016, Li2011}, the present work explicitly investigates the transport band gap of QSHI edge states using nonequilibrium Green’s function formalism. The transport gap is inversely proportional to the localization length \cite{Wakabayashi2009} and plays a crucial role in understanding electronic transport in disordered mesoscopic systems. We extract the explicit functional dependence of the transport gap on both the disorder and exchange proximity.
 The long-range order of the SSMs in combination with nonmagnetic disorder gives rise to the opening of the transport band gap, with an associated dependence on the wave vector of the spin spirals.

\section{Model}
Our goal is to study the impact of proximity from a 2D magnet on the electronic transport of a quantum spin Hall insulator. We show in Fig.~\ref{fig:fig1} the schematic of the device we are considering. It is a crossbar geometry van der Waals heterostructure formed by a QSHI with electrodes and a 2D SSM material on top of it. 1T$^{\prime}$-WTe$_2$ monolayer can serve as an illustrative example of a QSHI, while 1T-TaS$_2$ or NiI$_2$ monolayers can be potential candidates featuring the spin-spiral order. In this geometry, the exchange proximity effect gives rise to an exchange field in the QSHI, which is coupled with the helical edge states. 
For the sake of simplicity, the exchange proximity with magnetization in the \textit{xy} plane is included solely at the edges of the QSHI nanoribbon, where the electronic transport occurs. We will focus on studying the transport gap and the conductance measured with source and drain as depicted in Fig.~\ref{fig:fig1}. Nanoscale metallic electrodes can be deposited using modern fabrication techniques \cite{EBL2013, PVD_2022} or highly oriented pyrolitic graphite electrodes can be deterministically transferred on such structures.

The full Hamiltonian of the system takes the form $H = H_{QSH} + H_{J} + H_{dis}$, where $H_{QSH}$ and $H_{J}$ are Hamiltonians of a QSHI and spin-spiral exchange proximity, respectively, and $H_{dis}$ represents the nonmagnetic local disorder. The different terms are given by

\begin{equation*}
H_{QSH} = \sum_{\left\langle ij \right\rangle s}\left(t_{ij} c_{is}^\dagger c_{js}+h.c.\right) +
\end{equation*}
\begin{equation}
+ i \frac{\lambda_{SO}}{3\sqrt{3}} \sum_{\left\langle \left\langle ij \right\rangle\right\rangle s s'}\left([\mathbf{d}_{ik} \times \mathbf{d}_{kj}] \cdot \vec{\sigma}\;c_{is}^\dagger c_{js'}+h.c.\right),
\label{H_TI}
\end{equation}

\begin{equation}
H_{J} = \sum_{ss', \alpha \in \Xi}\left(\mathbf{J}_i\cdot \vec{\sigma}\right)_{ss'} c_{\alpha s}^\dagger c_{\alpha s'},
\label{H_SSM}
\end{equation}

\begin{equation}
    H_{dis} = \sum_{ is}W_i c_{is}^\dagger c_{is},
    \label{H_dis}
\end{equation}
where $c_{is}^\dagger$, $c_{is}$ are the creation and annihilation operators of the QSHI at site $i$ of the lattice and with specific spin projection, $t_{ij}$ are first neighbor hoppings \footnote{To incorporate edge effects, we take hopping to edge sites $t'=0.9t$, with $t$ hopping in bulk sites}.
The topological gap is controlled by
$\lambda_{SO} = 0.5t$, leading to the quantum spin Hall state, $\mathbf{{d}}_{ik},\mathbf{{d}}_{ik}$ denote vectors connecting sites $ik$, $kj$ \cite{KaneMele2005},
and $\vec{\sigma}$ consists of the Pauli matrices and acts on the spin degree of freedom of electrons.
The disorder in the system is included by random on-site energies 
$W_i \in [-W/2, W/2]$
as stemming from chemical disorder, where $W$ controls 
the strength of the random nonmagnetic disorder.  
Finally, $\mathbf{J}_i$ denotes the local exchange field induced by the spin-spiral material in the edge sites \cite{PhysRevB.94.144509, Chen2022spiral, PhysRevLett.133.236703}. We note that, while exchange coupling could have been included in the entire system, the existence of an edge is likely to make the QSH material more reactive at the edges \cite{Song_2021, Wang_2024} and hence such exchange coupling would dominate. The Appendix analyzes the scenario where the exchange coupling acts everywhere, showing analogous results. Thus, by $\Xi$ in Eq.~\eqref{H_SSM}, we denote sites that belong only to the edges of the QSHI nanoribbon.

The exchange profile $\mathbf{J}_i$ induced by the spiral can be written as follows:
\begin{equation}
\mathbf{J}_i = J_0 \left(\sin q (x_i - x_0), \cos{q  (x_i - x_0)}, \, 0 \right),
\label{magnetization}
\end{equation} 
where $J_0$ is the exchange coupling and $x_i$ is a coordinate of the site of the spiral. 
For the bulk spin-spiral material, the q vector $\mathbf q_0$
is a two-dimensional vector, where bulk magnetization is given by 
$\mathbf m (\mathbf r) 
= m(\cos(\mathbf q_0 \cdot \mathbf r), \sin(\mathbf q_0 \cdot \mathbf r), 0)$
for an in-plane spiral magnet. In terms of the helical modes of a QSHI, the effective exchange coupling on the edge sites depends on the projection of the bulk q vector along the edge boundary $\mathbf u$; such projection is denoted by $q = \mathbf q_0 \cdot \mathbf u$. It is worth noting that the effective $q$ can thus be controlled by changing the twist angle between the QSHI and the spin-spiral material, allowing one to explore different transport regimes with the same materials.
 We will quantify $q$ in units of $\frac{2\pi}{d}$, where 
$d$ is the lattice constant of the unit cell of the quantum spin Hall system.
 The value of $q$ defines the magnetic order that is projected on the edge sites: $q = 0$ and $q = 1/2$ correspond to ferromagnetic (FM) and antiferromagnetic (AFM) orders, respectively, whereas intermediate values give a generic noncollinear spiral state. 

\begin{figure}[t]
    \centering
    \includegraphics[width = 8.6cm]{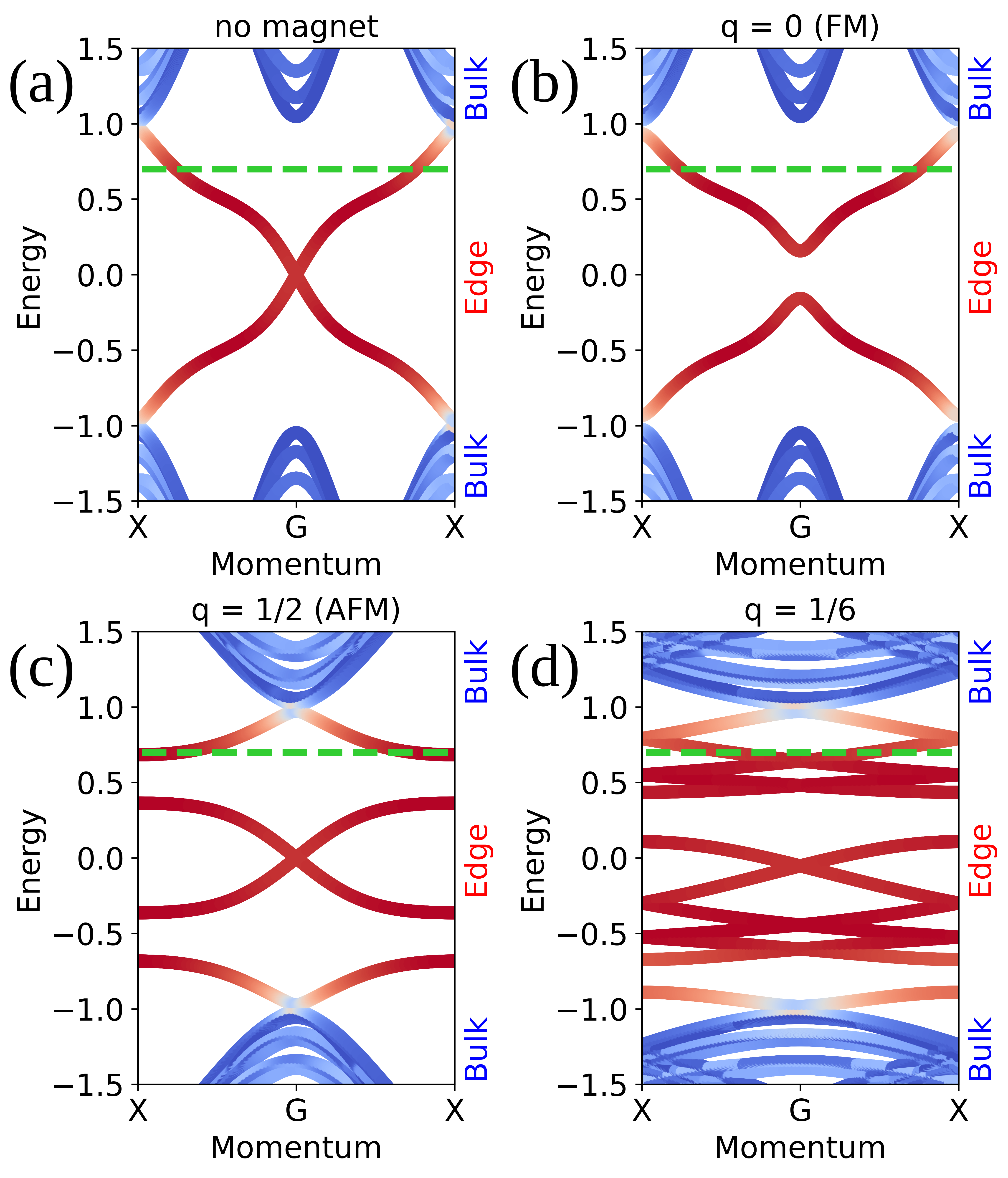} 
    \caption{ Bandstructure of the QSHI nanoribbon without (a) and with FM $q=0$ (b), AFM $q=1/2$ (c), and spin spiral exchange with $q = 1/6$ (d) in proximity. The green dashed line denotes Fermi level used in the transport simulations ($\mu = 0.7t$). The local exchange field causes the spectral gap opening in the helical modes, yet leaving edge states in a wide energy window inside the bulk gap.}
    \label{fig:fig2}
\end{figure}

We now discuss the influence of the spin-spiral exchange on the electronic transport of the topological edge states. In the presence of TRS breaking and nonmagnetic disorder, the edge states are expected to show a transport gap, despite not showing a spectral gap. This stems from the fact that, with a broken TRS, 1D modes will be localized due to Anderson localization, which occurs in our single-particle system at all energies, despite having a finite spectral weight. The band structure of the considered system, in the absence of the exchange field ($\mathbf{J}_i$ = 0), features topologically protected helical edge states that are gapless inside the bulk energy band gap [see Fig.~\ref{fig:fig2}(a)]. Turning on the exchange field breaks the TRS and opens a small spectral gap, yet leaves wide spectral regions gapless as shown in Figs.~\ref{fig:fig2}(b)-\ref{fig:fig2}(d) for the cases of the FM, AFM, and spin-spiral chains in proximity. 

\section{Transport gap of helical channels}
Topological edge states feature an absence of backscattering arising from the Kramer's partner of counterpropagating electrons \cite{KaneMele2005},
 enabling a 1D gas without Anderson localization as long as TRS is preserved. However, the presence of a nonmagnetic disorder in combination with a local magnetic field induced by the spirals in our system causes the appearance of backscattering and a decrease in conductance. Disorder effects would emerge from defects and impurities, which constitute an inevitable intrinsic property of these materials \cite{wu2016disorder}, or intentional doping \cite{Gupta2020, Wang2019}. We study the dependence of the emerging transport gap on the parameters of the projected spiral exchange field ($q$ and exchange coupling $J_0$) and the disorder strength $W$. The transport gap can be directly extracted from the channel length dependence of the conductance, which can be obtained from the Landauer formula given by \cite{Datta1997}

\begin{equation}
    I = \frac{2e}{h} \int dE (f(E, \mu_S)-f(E, \mu_D))\mathrm{Tr}(\G_c^\dagger \Gamma_D \G_c \Gamma_S),
    \label{Landauer}
\end{equation}
where $f(E, \mu)$ is the Fermi-distribution function; $\Gamma_{S, D} = i (\Sigma_{S, D} - \Sigma_{S, D}^\dagger)$, while $\Sigma_{S, D}$ are the self-energies and $\mu_{S, D}$ are chemical potentials of the source and drain electrodes, respectively; $\G_c$ is the Green's function of the coupled system. At zero temperature and for a small bias voltage applied across the electrodes, the conductance is obtained from Eq.~\eqref{Landauer} as follows:

\begin{equation}
    G = \frac{2e^2}{h} \mathrm{Tr}(\G_c^\dagger \Gamma_D \G_c \Gamma_S).
    \label{Landauer_conductance}
\end{equation}

\begin{figure}[t]
    \centering
    \includegraphics[width = 8.6cm]{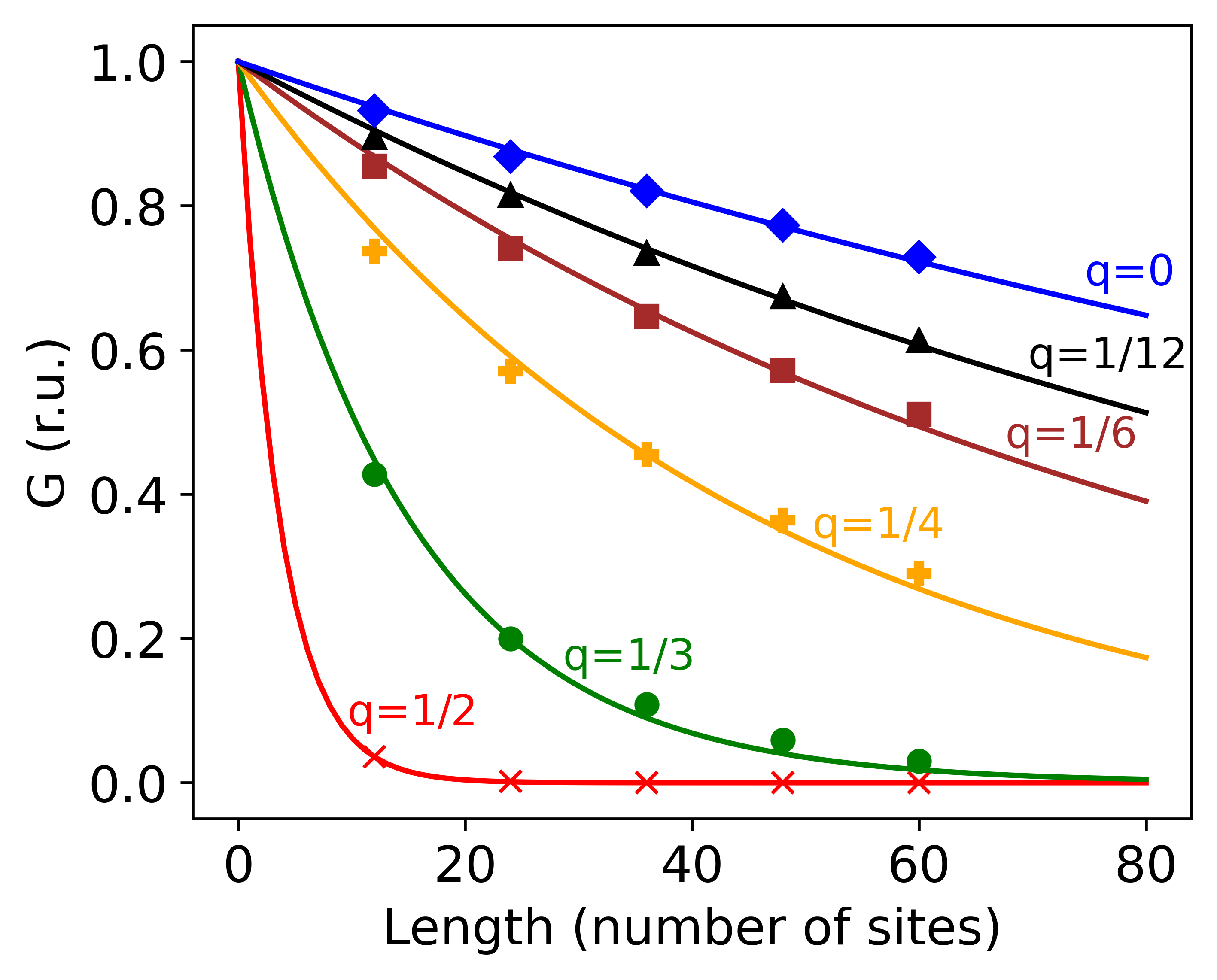} 
    \caption{Conductance as a function of the nanoribbon length for $W = 0.5t$ and different values of $q$. Solid lines show the fitting curves given by Eq.~\eqref{conductance}.}
    \label{fig:fig3}
\end{figure}

We performed transport calculations of the disordered QSHI system using the nonequilibrium Green's function formalism \cite{pyqula}. The nanoribbon width was chosen to reach the regime in which the hybridization between edge modes is suppressed \cite{Shafiei2022} so that interedge backscattering is not allowed.
 Using Eq.~\eqref{Landauer_conductance} we calculated the channel conductance $G$, which in our case is the function of the channel length, random disorder strength, spirals' exchange coupling, and $q$. The conductance at a chemical potential $\mu=0.7t$ corresponds to a value for which all band structures are gapless, thus allowing us to probe the transport properties of the helical edge states. In the experiment, the Fermi level in our or similar systems can be tuned by applying an electrical gate \cite{Fatemi2018}. Figure~\ref{fig:fig3} shows the dependence of the conductance on the channel length for $W = 0.5t$ and different values of $q$. We averaged the conductance over 1000 realizations of disorder in Eq.~\eqref{H_dis}. The emergence of the transport gap due to Anderson localization leads to exponential length-dependent conductance of the helical states for any values of $q$ in the form \cite{Abrahams1979}
\begin{equation}
    G = G_{0} e^{-\Delta L},
    \label{conductance}
\end{equation}
where $G_0 = 2e^2/h$ is the conductance quantum, $L$ is the channel length, and $\Delta$ is the transport gap \cite{Wakabayashi2009}. We further extract $\Delta$ for different $q$ values as a fitting parameter by the approximation of the conductance given by Eq.~\eqref{conductance} to the simulation data points. We emphasize that, to obtain backscattering and, as a result, exponential decay of the conductance with the channel length, two constituents are required: an exchange field and a random disorder. Without either of the two constituents, there is no backscattering in topologically protected helical channels and the conductance decay does not emerge ($G = G_0$ and $\Delta = 0$).

\begin{figure}[t]
    \centering
    \includegraphics[width = 8.6cm]{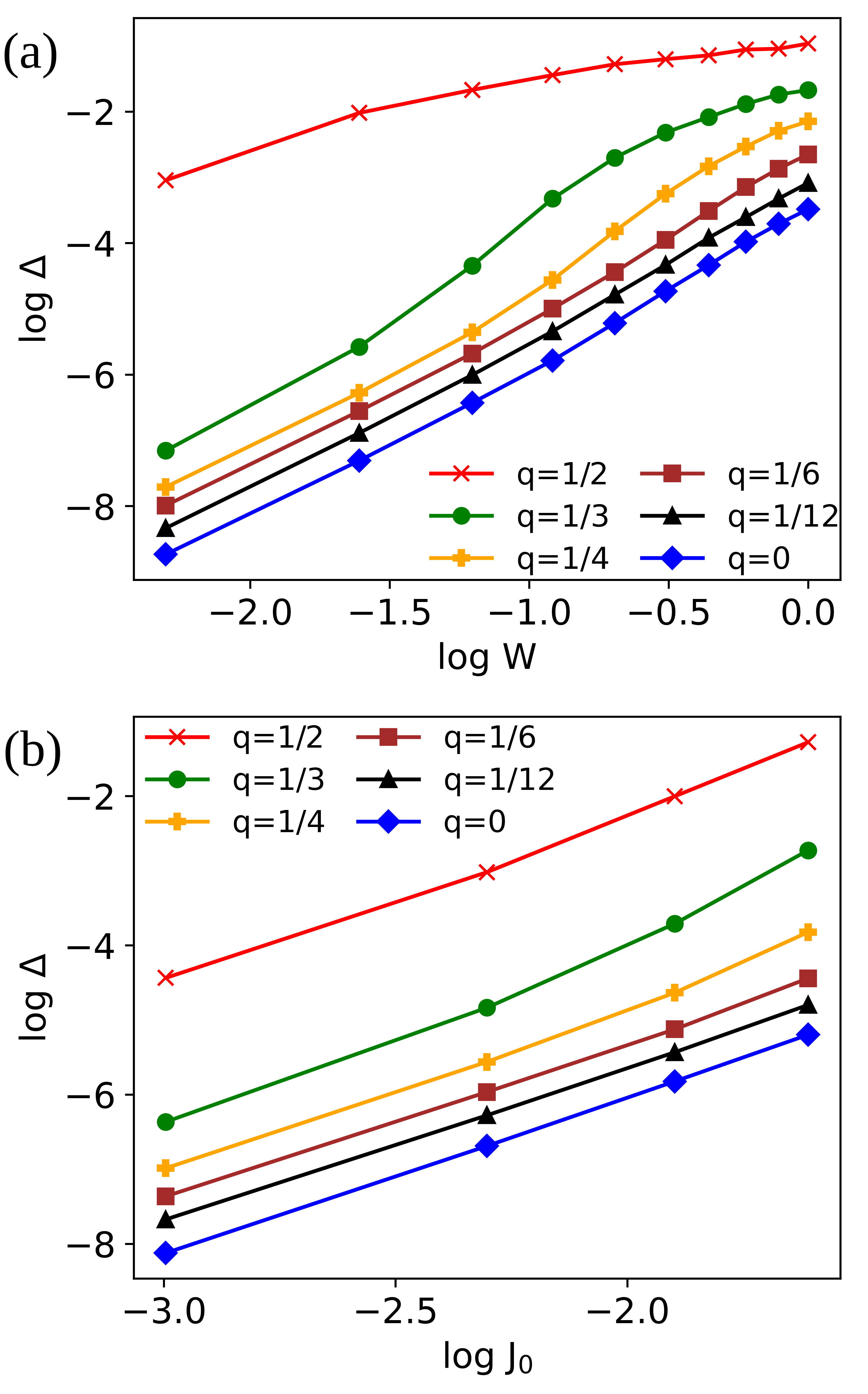} 
    \caption{Log-log dependence of the transport gap $\Delta$ vs random disorder strength $W$ (a) and exchange coupling $J_0$ (b) for different values of $q$. The linear behavior of the
    log-log plot reflects the power-law dependence of the transport gap as a function of $J_0$ and $W$.}
    \label{fig:fig4}
\end{figure}

\begin{figure}[t]
    \centering
    \includegraphics[width = 8.6cm]{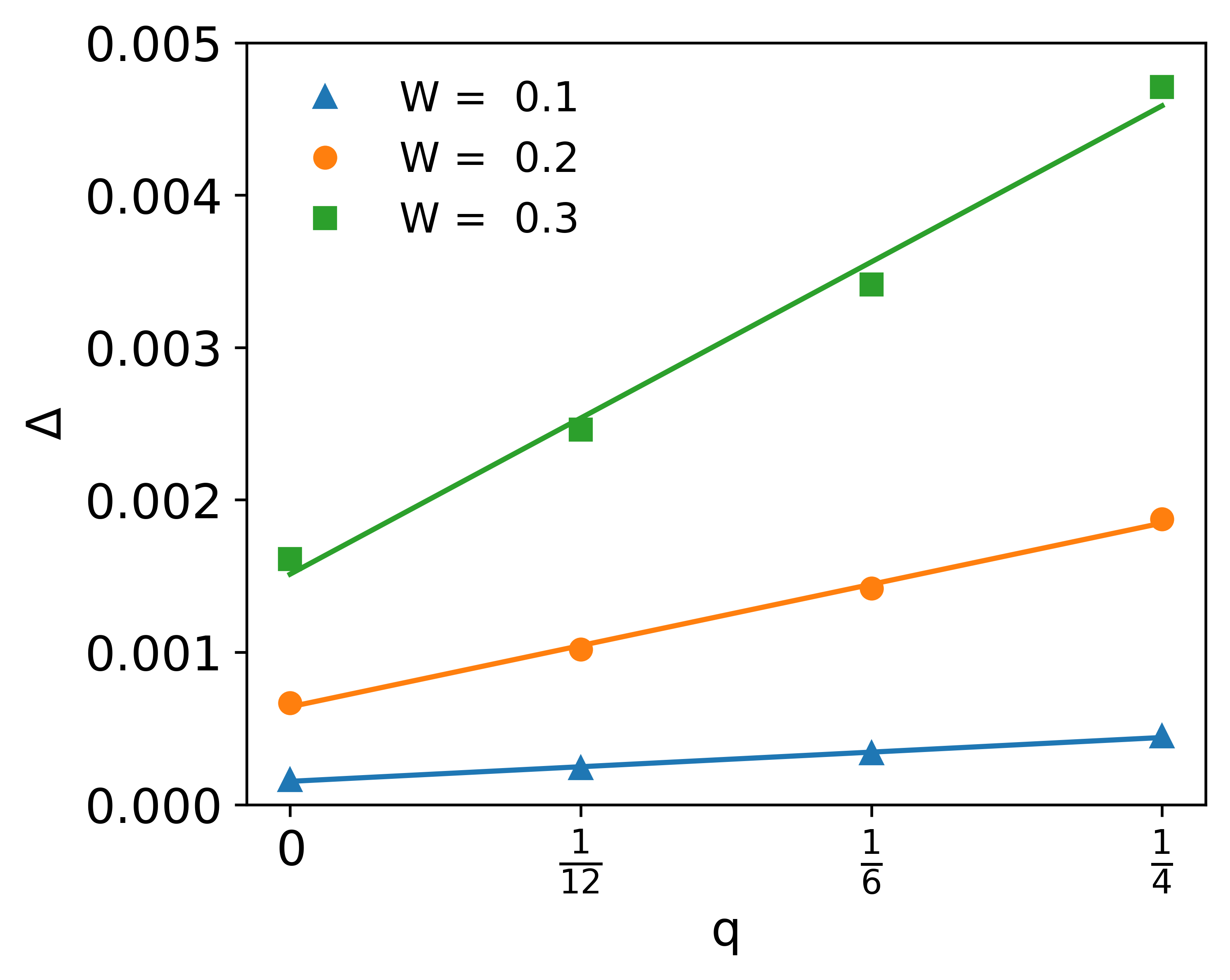} 
    \caption{Dependence of the transport gap $\Delta$ vs $q$ for different values of the disorder strength. A linear dependence of the transport gap is observed for small q vectors.}
    \label{fig:fig5}
\end{figure}

Figures~\ref{fig:fig4}(a) and \ref{fig:fig4}(b) show the obtained log-log dependences of the transport gap as a function of the disorder strength and exchange coupling, respectively. Both plots reveal a linear dependence for sufficiently small values of the parameters, which corresponds to a quadratic power-law dependence of the transport gap.
 At the same time, the dependence of the transport gap on $q$ is shown to be linear for small values of $q$ as can be seen in Fig.~\ref{fig:fig5} plotted for different values of the disorder strength ($W \in [0.1t,0.3t]$). This dependence allows us to qualitatively obtain the q vector of the magnet by measuring transport at different twist angles between the QSHI and the SSM \cite{Yang2020}.

The results presented in Figs.~\ref{fig:fig4}(a) and \ref{fig:fig4}(b) and Fig.~{\ref{fig:fig5}} allow us to extract the dependence of the transport gap for small values of $J_0$, $W$, and $q$ as

\begin{equation}
    \Delta \propto J_0^2 W^2 (1+\gamma \cdot |q|),
    \label{gap_formula}
\end{equation}
where powers of $J_0$ and $W$ and also factor $\gamma$ can be obtained as fitting parameters. 
The integer power dependence suggests that both mechanisms of the gap opening arise from perturbation theory in disorder and an exchange field from SSMs. In the weak coupling regime considered here ($J_0 \ll t$, $W \ll t$), the power-law dependence can be rationalized from perturbation theory with the disordered Green's functions. Thus, both disorder strength and exchange coupling are instrumental in leading to an exponential localization of the states and quenching transport in the helical channels. In particular, since for quantum spin liquid the expectation value of the exchange field vanishes, no backscattering will be expected from the mechanism described in our work. It is nonetheless worth mentioning that, for a quantum spin liquid, a transport gap may still appear due to an emerging heavy-fermion gap due to Kondo coupling \cite{Wirth2016, Posey2024, Vano2021,2024arXiv240917202L, PhysRevResearch.6.023227}. Such a Kondo gap will have a very strong temperature dependence \cite{Wirth2016} and thus it can be distinguished from a transport gap stemming from backscattering.

Since various effects can influence conductance in real experiments, it is important to emphasize that the theoretical results presented in this paper are obtained under specific assumptions that allow us to isolate the subject of our study. In particular, to obtain conductance changes induced purely by changing the q vector, we assume a constant exchange coupling $J_{0}$. Moreover, we neglect strong electron-electron interactions and many-body effects, including, among others, Kondo screening, which enables us to consider Anderson localization within a single-particle framework. It is finally worth pointing out that our results assume a perfectly ordered spin spiral and, therefore, our analysis is valid at temperatures below the critical temperature of the spin-spiral material.

\section{Conclusions}
In this paper, we have analyzed the influence of the proximity effect driven by a spin-spiral magnet on the electronic transport of topological edge states in a QSHI with a nonmagnetic disorder. The local exchange field created by spirals combined with nonmagnetic disorder results in the emergence of backscattering and the emergence of a transport gap for helical edge states.  We have obtained an explicit functional dependence for the transport gap that reveals the power-law dependences on the exchange coupling, disorder strength, and a linear dependence on the absolute value of the wave vector of the spin spirals. The conductance drop driven by the combination of these mechanisms serves as an indicator of a noncollinear magnetic state. Our findings show that the helical states of a QSHI may serve as a potential sensor of spiral magnetism in low-dimensional materials. The experimentally accessible low-temperature conductance measurements \cite{konig2007, Lee2018} in our system can help to characterize and distinguish spin-spiral states from other phases, in particular, quantum spin liquids that do not break TRS. Our strategy can be effective for understanding the magnetic nature of some 2D materials, for example, those that support frustrated magnetism like 1T-TaS$_{2}$ monolayer, or multiferroic materials like NiI$_2$ monolayer. These results suggest electronic transport of topological edge states in the QSHI/SSM van der Waals heterostructures as a promising method to probe noncollinear magnetization at the nanoscale by electrical means.

\section*{Data availability}
The data that support the findings of this article are openly
available \cite{nigmatulin_2025_15755975}.
 
\begin{acknowledgments}
 This work was supported by the Research Council of  Finland Flagship Programme (Grant No. 320167, PREIN), the EU H2020-MSCA-RISE-872049 (IPN-Bio), ERC (Grant No. 834742), the Academy of Finland Projects No. 331342 and No. 358088, the Finnish Quantum Flagship, the School of Electrical Engineering (Aalto University), and the Nokia Foundation. We acknowledge the computational resources provided by the Aalto Science-IT project.
\end{acknowledgments}
\vspace{12pt}

\appendix*
\section{Conductance in the presence of bulk magnetization}
\begin{figure}[ht]
    \centering
    \includegraphics[width=8.6cm]{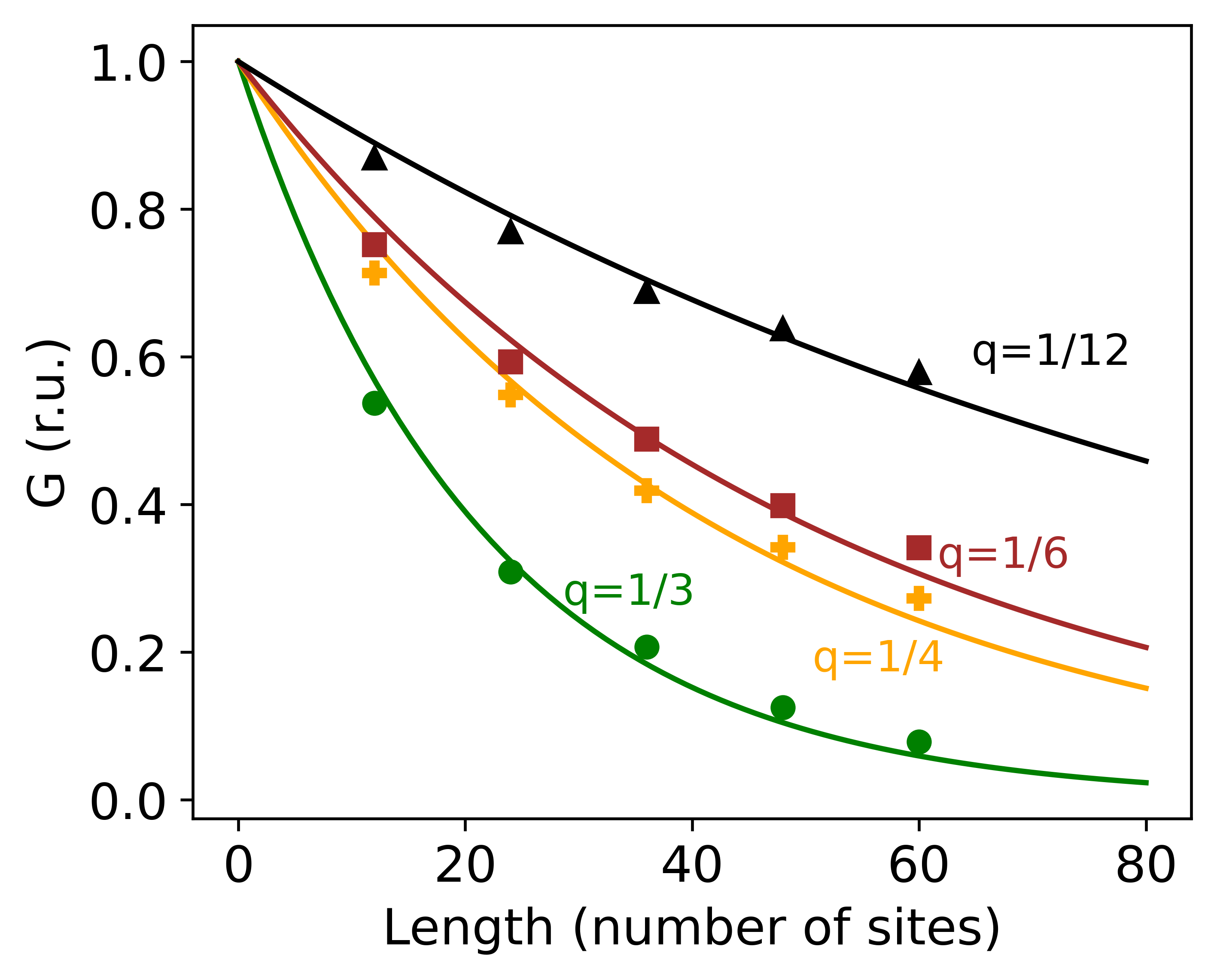}
    \caption{Conductance as a function of the nanoribbon length for $W = 0.5t$ and different values of $q$ with the presence of the bulk magnetization of the entire 2D SSM.}
    \label{fig:bulk_magnetization}
\end{figure}

The bulk magnetization can be included in our system by adding the exchange coupling given by Eq.~\eqref{H_SSM} to every site of the QSHI. However, since the conductance is provided by the edge states, a dominant effect of the exchange coupling 
occurs at the edges of the QSHI for energies inside the topological gap. We present simulation results for 2D SSMs with different spin-spiral q vectors in Fig.~\ref{fig:bulk_magnetization}, with exchange included everywhere. As one can see,
besides a small renormalization of the decay in comparison with Fig.~\ref{fig:fig3}, the general behavior of the conductance remains the same as for the case of the solely edge magnetization described in our manuscript.

\label{appendix}
\bibliography{apssamp}

\begin{thebibliography}{67}%
\makeatletter
\providecommand \@ifxundefined [1]{%
 \@ifx{#1\undefined}
}%
\providecommand \@ifnum [1]{%
 \ifnum #1\expandafter \@firstoftwo
 \else \expandafter \@secondoftwo
 \fi
}%
\providecommand \@ifx [1]{%
 \ifx #1\expandafter \@firstoftwo
 \else \expandafter \@secondoftwo
 \fi
}%
\providecommand \natexlab [1]{#1}%
\providecommand \enquote  [1]{``#1''}%
\providecommand \bibnamefont  [1]{#1}%
\providecommand \bibfnamefont [1]{#1}%
\providecommand \citenamefont [1]{#1}%
\providecommand \href@noop [0]{\@secondoftwo}%
\providecommand \href [0]{\begingroup \@sanitize@url \@href}%
\providecommand \@href[1]{\@@startlink{#1}\@@href}%
\providecommand \@@href[1]{\endgroup#1\@@endlink}%
\providecommand \@sanitize@url [0]{\catcode `\\12\catcode `\$12\catcode `\&12\catcode `\#12\catcode `\^12\catcode `\_12\catcode `\%12\relax}%
\providecommand \@@startlink[1]{}%
\providecommand \@@endlink[0]{}%
\providecommand \url  [0]{\begingroup\@sanitize@url \@url }%
\providecommand \@url [1]{\endgroup\@href {#1}{\urlprefix }}%
\providecommand \urlprefix  [0]{URL }%
\providecommand \Eprint [0]{\href }%
\providecommand \doibase [0]{https://doi.org/}%
\providecommand \selectlanguage [0]{\@gobble}%
\providecommand \bibinfo  [0]{\@secondoftwo}%
\providecommand \bibfield  [0]{\@secondoftwo}%
\providecommand \translation [1]{[#1]}%
\providecommand \BibitemOpen [0]{}%
\providecommand \bibitemStop [0]{}%
\providecommand \bibitemNoStop [0]{.\EOS\space}%
\providecommand \EOS [0]{\spacefactor3000\relax}%
\providecommand \BibitemShut  [1]{\csname bibitem#1\endcsname}%
\let\auto@bib@innerbib\@empty
\bibitem [{\citenamefont {Bousquet}\ and\ \citenamefont {Cano}(2016)}]{bousquet2016}%
  \BibitemOpen
  \bibfield  {author} {\bibinfo {author} {\bibfnamefont {E.}~\bibnamefont {Bousquet}}\ and\ \bibinfo {author} {\bibfnamefont {A.}~\bibnamefont {Cano}},\ }\bibfield  {title} {\bibinfo {title} {Non-collinear magnetism in multiferroic perovskites},\ }\href {https://doi.org/10.1088/0953-8984/28/12/123001} {\bibfield  {journal} {\bibinfo  {journal} {J. Phys. Condens. Matter}\ }\textbf {\bibinfo {volume} {28}},\ \bibinfo {pages} {123001} (\bibinfo {year} {2016})}\BibitemShut {NoStop}%
\bibitem [{\citenamefont {Steinbrecher}\ \emph {et~al.}(2018)\citenamefont {Steinbrecher}, \citenamefont {Rausch}, \citenamefont {That}, \citenamefont {Hermenau}, \citenamefont {Khajetoorians}, \citenamefont {Potthoff}, \citenamefont {Wiesendanger},\ and\ \citenamefont {Wiebe}}]{steinbrecher2018}%
  \BibitemOpen
  \bibfield  {author} {\bibinfo {author} {\bibfnamefont {M.}~\bibnamefont {Steinbrecher}}, \bibinfo {author} {\bibfnamefont {R.}~\bibnamefont {Rausch}}, \bibinfo {author} {\bibfnamefont {K.~T.}\ \bibnamefont {That}}, \bibinfo {author} {\bibfnamefont {J.}~\bibnamefont {Hermenau}}, \bibinfo {author} {\bibfnamefont {A.~A.}\ \bibnamefont {Khajetoorians}}, \bibinfo {author} {\bibfnamefont {M.}~\bibnamefont {Potthoff}}, \bibinfo {author} {\bibfnamefont {R.}~\bibnamefont {Wiesendanger}},\ and\ \bibinfo {author} {\bibfnamefont {J.}~\bibnamefont {Wiebe}},\ }\bibfield  {title} {\bibinfo {title} {Non-collinear spin states in bottom-up fabricated atomic chains},\ }\href {https://doi.org/10.1038/s41467-018-05364-5} {\bibfield  {journal} {\bibinfo  {journal} {Nat. Commun.}\ }\textbf {\bibinfo {volume} {9}},\ \bibinfo {pages} {2853} (\bibinfo {year} {2018})}\BibitemShut {NoStop}%
\bibitem [{\citenamefont {von Bergmann}\ \emph {et~al.}(2014)\citenamefont {von Bergmann}, \citenamefont {Kubetzka}, \citenamefont {Pietzsch},\ and\ \citenamefont {Wiesendanger}}]{Bergmann2014}%
  \BibitemOpen
  \bibfield  {author} {\bibinfo {author} {\bibfnamefont {K.}~\bibnamefont {von Bergmann}}, \bibinfo {author} {\bibfnamefont {A.}~\bibnamefont {Kubetzka}}, \bibinfo {author} {\bibfnamefont {O.}~\bibnamefont {Pietzsch}},\ and\ \bibinfo {author} {\bibfnamefont {R.}~\bibnamefont {Wiesendanger}},\ }\bibfield  {title} {\bibinfo {title} {Interface-induced chiral domain walls, spin spirals and skyrmions revealed by spin-polarized scanning tunneling microscopy},\ }\href {https://doi.org/10.1088/0953-8984/26/39/394002} {\bibfield  {journal} {\bibinfo  {journal} {J. Phys. Condens. Matter}\ }\textbf {\bibinfo {volume} {26}},\ \bibinfo {pages} {394002} (\bibinfo {year} {2014})}\BibitemShut {NoStop}%
\bibitem [{\citenamefont {Menzel}\ \emph {et~al.}(2012)\citenamefont {Menzel}, \citenamefont {Mokrousov}, \citenamefont {Wieser}, \citenamefont {Bickel}, \citenamefont {Vedmedenko}, \citenamefont {Bl\"ugel}, \citenamefont {Heinze}, \citenamefont {von Bergmann}, \citenamefont {Kubetzka},\ and\ \citenamefont {Wiesendanger}}]{Menzel2012}%
  \BibitemOpen
  \bibfield  {author} {\bibinfo {author} {\bibfnamefont {M.}~\bibnamefont {Menzel}}, \bibinfo {author} {\bibfnamefont {Y.}~\bibnamefont {Mokrousov}}, \bibinfo {author} {\bibfnamefont {R.}~\bibnamefont {Wieser}}, \bibinfo {author} {\bibfnamefont {J.~E.}\ \bibnamefont {Bickel}}, \bibinfo {author} {\bibfnamefont {E.}~\bibnamefont {Vedmedenko}}, \bibinfo {author} {\bibfnamefont {S.}~\bibnamefont {Bl\"ugel}}, \bibinfo {author} {\bibfnamefont {S.}~\bibnamefont {Heinze}}, \bibinfo {author} {\bibfnamefont {K.}~\bibnamefont {von Bergmann}}, \bibinfo {author} {\bibfnamefont {A.}~\bibnamefont {Kubetzka}},\ and\ \bibinfo {author} {\bibfnamefont {R.}~\bibnamefont {Wiesendanger}},\ }\bibfield  {title} {\bibinfo {title} {Information transfer by vector spin chirality in finite magnetic chains},\ }\href {https://doi.org/10.1103/PhysRevLett.108.197204} {\bibfield  {journal} {\bibinfo  {journal} {Phys. Rev. Lett.}\ }\textbf {\bibinfo {volume} {108}},\ \bibinfo {pages} {197204} (\bibinfo {year} {2012})}\BibitemShut {NoStop}%
\bibitem [{\citenamefont {Rimmler}\ \emph {et~al.}(2024)\citenamefont {Rimmler}, \citenamefont {Pal},\ and\ \citenamefont {Parkin}}]{rimmler2024}%
  \BibitemOpen
  \bibfield  {author} {\bibinfo {author} {\bibfnamefont {B.~H.}\ \bibnamefont {Rimmler}}, \bibinfo {author} {\bibfnamefont {B.}~\bibnamefont {Pal}},\ and\ \bibinfo {author} {\bibfnamefont {S.~S.}\ \bibnamefont {Parkin}},\ }\bibfield  {title} {\bibinfo {title} {Non-collinear antiferromagnetic spintronics},\ }\href {https://doi.org/10.1038/s41578-024-00706-w} {\bibfield  {journal} {\bibinfo  {journal} {Nat. Rev. Mater.}\ }\textbf {\bibinfo {volume} {10}},\ \bibinfo {pages} {109} (\bibinfo {year} {2024})}\BibitemShut {NoStop}%
\bibitem [{\citenamefont {Nigmatulin}\ \emph {et~al.}(2021)\citenamefont {Nigmatulin}, \citenamefont {Shelykh},\ and\ \citenamefont {Iorsh}}]{Nigmatulin2021}%
  \BibitemOpen
  \bibfield  {author} {\bibinfo {author} {\bibfnamefont {F.~O.}\ \bibnamefont {Nigmatulin}}, \bibinfo {author} {\bibfnamefont {I.~A.}\ \bibnamefont {Shelykh}},\ and\ \bibinfo {author} {\bibfnamefont {I.~V.}\ \bibnamefont {Iorsh}},\ }\bibfield  {title} {\bibinfo {title} {Quantum spin compass models in two-dimensional electronic topological metasurfaces},\ }\href {https://doi.org/10.1103/PhysRevResearch.3.043016} {\bibfield  {journal} {\bibinfo  {journal} {Phys. Rev. Res.}\ }\textbf {\bibinfo {volume} {3}},\ \bibinfo {pages} {043016} (\bibinfo {year} {2021})}\BibitemShut {NoStop}%
\bibitem [{\citenamefont {Tokura}\ and\ \citenamefont {Seki}(2010)}]{Tokura2010}%
  \BibitemOpen
  \bibfield  {author} {\bibinfo {author} {\bibfnamefont {Y.}~\bibnamefont {Tokura}}\ and\ \bibinfo {author} {\bibfnamefont {S.}~\bibnamefont {Seki}},\ }\bibfield  {title} {\bibinfo {title} {Multiferroics with spiral spin orders},\ }\href {https://onlinelibrary.wiley.com/doi/abs/10.1002/adma.200901961} {\bibfield  {journal} {\bibinfo  {journal} {Adv. Mater.}\ }\textbf {\bibinfo {volume} {22}},\ \bibinfo {pages} {1554} (\bibinfo {year} {2010})}\BibitemShut {NoStop}%
\bibitem [{\citenamefont {Song}\ \emph {et~al.}(2022)\citenamefont {Song}, \citenamefont {Occhialini}, \citenamefont {Erge\c{c}en}, \citenamefont {Ilyas}, \citenamefont {Amoroso}, \citenamefont {Barone}, \citenamefont {Kapeghian}, \citenamefont {Watanabe}, \citenamefont {Taniguchi}, \citenamefont {Botana}, \citenamefont {Picozzi}, \citenamefont {Gedik},\ and\ \citenamefont {Comin}}]{Song2022}%
  \BibitemOpen
  \bibfield  {author} {\bibinfo {author} {\bibfnamefont {Q.}~\bibnamefont {Song}}, \bibinfo {author} {\bibfnamefont {C.~A.}\ \bibnamefont {Occhialini}}, \bibinfo {author} {\bibfnamefont {E.}~\bibnamefont {Erge\c{c}en}}, \bibinfo {author} {\bibfnamefont {B.}~\bibnamefont {Ilyas}}, \bibinfo {author} {\bibfnamefont {D.}~\bibnamefont {Amoroso}}, \bibinfo {author} {\bibfnamefont {P.}~\bibnamefont {Barone}}, \bibinfo {author} {\bibfnamefont {J.}~\bibnamefont {Kapeghian}}, \bibinfo {author} {\bibfnamefont {K.}~\bibnamefont {Watanabe}}, \bibinfo {author} {\bibfnamefont {T.}~\bibnamefont {Taniguchi}}, \bibinfo {author} {\bibfnamefont {A.~S.}\ \bibnamefont {Botana}}, \bibinfo {author} {\bibfnamefont {S.}~\bibnamefont {Picozzi}}, \bibinfo {author} {\bibfnamefont {N.}~\bibnamefont {Gedik}},\ and\ \bibinfo {author} {\bibfnamefont {R.}~\bibnamefont {Comin}},\ }\bibfield  {title} {\bibinfo {title} {Evidence for a single-layer van der {W}aals multiferroic},\ }\href {https://doi.org/10.1038/s41586-021-04337-x} {\bibfield
  {journal} {\bibinfo  {journal} {Nature}\ }\textbf {\bibinfo {volume} {602}},\ \bibinfo {pages} {601–605} (\bibinfo {year} {2022})}\BibitemShut {NoStop}%
\bibitem [{\citenamefont {Tokura}\ and\ \citenamefont {Kanazawa}(2021)}]{tokura2021}%
  \BibitemOpen
  \bibfield  {author} {\bibinfo {author} {\bibfnamefont {Y.}~\bibnamefont {Tokura}}\ and\ \bibinfo {author} {\bibfnamefont {N.}~\bibnamefont {Kanazawa}},\ }\bibfield  {title} {\bibinfo {title} {Magnetic skyrmion materials},\ }\href {https://doi.org/10.1021/acs.chemrev.0c00297} {\bibfield  {journal} {\bibinfo  {journal} {Chem. Rev.}\ }\textbf {\bibinfo {volume} {121}},\ \bibinfo {pages} {2857} (\bibinfo {year} {2021})}\BibitemShut {NoStop}%
\bibitem [{\citenamefont {Wuhrer}\ \emph {et~al.}(2023)\citenamefont {Wuhrer}, \citenamefont {R\'ozsa}, \citenamefont {Nowak},\ and\ \citenamefont {Belzig}}]{Wuhrer2023}%
  \BibitemOpen
  \bibfield  {author} {\bibinfo {author} {\bibfnamefont {D.}~\bibnamefont {Wuhrer}}, \bibinfo {author} {\bibfnamefont {L.}~\bibnamefont {R\'ozsa}}, \bibinfo {author} {\bibfnamefont {U.}~\bibnamefont {Nowak}},\ and\ \bibinfo {author} {\bibfnamefont {W.}~\bibnamefont {Belzig}},\ }\bibfield  {title} {\bibinfo {title} {Magnon squeezing in conical spin spirals},\ }\href {https://doi.org/10.1103/PhysRevResearch.5.043124} {\bibfield  {journal} {\bibinfo  {journal} {Phys. Rev. Res.}\ }\textbf {\bibinfo {volume} {5}},\ \bibinfo {pages} {043124} (\bibinfo {year} {2023})}\BibitemShut {NoStop}%
\bibitem [{\citenamefont {Fert}\ \emph {et~al.}(2017)\citenamefont {Fert}, \citenamefont {Reyren},\ and\ \citenamefont {Cros}}]{Fert2017}%
  \BibitemOpen
  \bibfield  {author} {\bibinfo {author} {\bibfnamefont {A.}~\bibnamefont {Fert}}, \bibinfo {author} {\bibfnamefont {N.}~\bibnamefont {Reyren}},\ and\ \bibinfo {author} {\bibfnamefont {V.}~\bibnamefont {Cros}},\ }\bibfield  {title} {\bibinfo {title} {Magnetic skyrmions: advances in physics and potential applications},\ }\href {https://doi.org/10.1038/natrevmats.2017.31} {\bibfield  {journal} {\bibinfo  {journal} {Nat. Rev. Mater.}\ }\textbf {\bibinfo {volume} {2}},\ \bibinfo {pages} {1} (\bibinfo {year} {2017})}\BibitemShut {NoStop}%
\bibitem [{\citenamefont {Furrer}\ \emph {et~al.}(2009)\citenamefont {Furrer}, \citenamefont {Mesot},\ and\ \citenamefont {Str{\"a}ssle}}]{furrer2009}%
  \BibitemOpen
  \bibfield  {author} {\bibinfo {author} {\bibfnamefont {A.}~\bibnamefont {Furrer}}, \bibinfo {author} {\bibfnamefont {J.~F.}\ \bibnamefont {Mesot}},\ and\ \bibinfo {author} {\bibfnamefont {T.}~\bibnamefont {Str{\"a}ssle}},\ }\href@noop {} {\emph {\bibinfo {title} {Neutron scattering in condensed matter physics}}},\ Vol.~\bibinfo {volume} {4}\ (\bibinfo  {publisher} {World Sci.},\ \bibinfo {address} {Singapore},\ \bibinfo {year} {2009})\BibitemShut {NoStop}%
\bibitem [{\citenamefont {M{\"u}hlbauer}\ \emph {et~al.}(2019)\citenamefont {M{\"u}hlbauer}, \citenamefont {Honecker}, \citenamefont {P{\'e}rigo}, \citenamefont {Bergner}, \citenamefont {Disch}, \citenamefont {Heinemann}, \citenamefont {Erokhin}, \citenamefont {Berkov}, \citenamefont {Leighton}, \citenamefont {Eskildsen},\ and\ \citenamefont {Michels}}]{muhlbauer2019}%
  \BibitemOpen
  \bibfield  {author} {\bibinfo {author} {\bibfnamefont {S.}~\bibnamefont {M{\"u}hlbauer}}, \bibinfo {author} {\bibfnamefont {D.}~\bibnamefont {Honecker}}, \bibinfo {author} {\bibfnamefont {{\'E}.~A.}\ \bibnamefont {P{\'e}rigo}}, \bibinfo {author} {\bibfnamefont {F.}~\bibnamefont {Bergner}}, \bibinfo {author} {\bibfnamefont {S.}~\bibnamefont {Disch}}, \bibinfo {author} {\bibfnamefont {A.}~\bibnamefont {Heinemann}}, \bibinfo {author} {\bibfnamefont {S.}~\bibnamefont {Erokhin}}, \bibinfo {author} {\bibfnamefont {D.}~\bibnamefont {Berkov}}, \bibinfo {author} {\bibfnamefont {C.}~\bibnamefont {Leighton}}, \bibinfo {author} {\bibfnamefont {M.~R.}\ \bibnamefont {Eskildsen}},\ and\ \bibinfo {author} {\bibfnamefont {A.}~\bibnamefont {Michels}},\ }\bibfield  {title} {\bibinfo {title} {Magnetic small-angle neutron scattering},\ }\href {https://doi.org/10.1103/RevModPhys.91.015004} {\bibfield  {journal} {\bibinfo  {journal} {Rev. Mod. Phys.}\ }\textbf {\bibinfo {volume} {91}},\ \bibinfo {pages} {015004} (\bibinfo {year}
  {2019})}\BibitemShut {NoStop}%
\bibitem [{\citenamefont {Bode}\ \emph {et~al.}(2007)\citenamefont {Bode}, \citenamefont {Heide}, \citenamefont {Von~Bergmann}, \citenamefont {Ferriani}, \citenamefont {Heinze}, \citenamefont {Bihlmayer}, \citenamefont {Kubetzka}, \citenamefont {Pietzsch}, \citenamefont {Bl{\"u}gel},\ and\ \citenamefont {Wiesendanger}}]{bode2007}%
  \BibitemOpen
  \bibfield  {author} {\bibinfo {author} {\bibfnamefont {M.}~\bibnamefont {Bode}}, \bibinfo {author} {\bibfnamefont {M.}~\bibnamefont {Heide}}, \bibinfo {author} {\bibfnamefont {K.}~\bibnamefont {Von~Bergmann}}, \bibinfo {author} {\bibfnamefont {P.}~\bibnamefont {Ferriani}}, \bibinfo {author} {\bibfnamefont {S.}~\bibnamefont {Heinze}}, \bibinfo {author} {\bibfnamefont {G.}~\bibnamefont {Bihlmayer}}, \bibinfo {author} {\bibfnamefont {A.}~\bibnamefont {Kubetzka}}, \bibinfo {author} {\bibfnamefont {O.}~\bibnamefont {Pietzsch}}, \bibinfo {author} {\bibfnamefont {S.}~\bibnamefont {Bl{\"u}gel}},\ and\ \bibinfo {author} {\bibfnamefont {R.}~\bibnamefont {Wiesendanger}},\ }\bibfield  {title} {\bibinfo {title} {Chiral magnetic order at surfaces driven by inversion asymmetry},\ }\href {https://doi.org/10.1038/nature05802} {\bibfield  {journal} {\bibinfo  {journal} {Nature}\ }\textbf {\bibinfo {volume} {447}},\ \bibinfo {pages} {190} (\bibinfo {year} {2007})}\BibitemShut {NoStop}%
\bibitem [{\citenamefont {Fumega}\ and\ \citenamefont {Lado}(2022)}]{Fumega2022}%
  \BibitemOpen
  \bibfield  {author} {\bibinfo {author} {\bibfnamefont {A.~O.}\ \bibnamefont {Fumega}}\ and\ \bibinfo {author} {\bibfnamefont {J.~L.}\ \bibnamefont {Lado}},\ }\bibfield  {title} {\bibinfo {title} {Microscopic origin of multiferroic order in monolayer {N}i{I}$_2$},\ }\href {https://doi.org/10.1088/2053-1583/ac4e9d} {\bibfield  {journal} {\bibinfo  {journal} {2D Mater.}\ }\textbf {\bibinfo {volume} {9}},\ \bibinfo {pages} {025010} (\bibinfo {year} {2022})}\BibitemShut {NoStop}%
\bibitem [{\citenamefont {Li}\ \emph {et~al.}(2023)\citenamefont {Li}, \citenamefont {Xu}, \citenamefont {Liu}, \citenamefont {Li}, \citenamefont {Bellaiche},\ and\ \citenamefont {Xiang}}]{PhysRevLett.131.036701}%
  \BibitemOpen
  \bibfield  {author} {\bibinfo {author} {\bibfnamefont {X.}~\bibnamefont {Li}}, \bibinfo {author} {\bibfnamefont {C.}~\bibnamefont {Xu}}, \bibinfo {author} {\bibfnamefont {B.}~\bibnamefont {Liu}}, \bibinfo {author} {\bibfnamefont {X.}~\bibnamefont {Li}}, \bibinfo {author} {\bibfnamefont {L.}~\bibnamefont {Bellaiche}},\ and\ \bibinfo {author} {\bibfnamefont {H.}~\bibnamefont {Xiang}},\ }\bibfield  {title} {\bibinfo {title} {Realistic spin model for multiferroic {N}i{I}$_{2}$},\ }\href {https://doi.org/10.1103/PhysRevLett.131.036701} {\bibfield  {journal} {\bibinfo  {journal} {Phys. Rev. Lett.}\ }\textbf {\bibinfo {volume} {131}},\ \bibinfo {pages} {036701} (\bibinfo {year} {2023})}\BibitemShut {NoStop}%
\bibitem [{\citenamefont {Riedl}\ \emph {et~al.}(2022)\citenamefont {Riedl}, \citenamefont {Amoroso}, \citenamefont {Backes}, \citenamefont {Razpopov}, \citenamefont {Nguyen}, \citenamefont {Yamauchi}, \citenamefont {Barone}, \citenamefont {Winter}, \citenamefont {Picozzi},\ and\ \citenamefont {Valent\'{\i}}}]{PhysRevB.106.035156}%
  \BibitemOpen
  \bibfield  {author} {\bibinfo {author} {\bibfnamefont {K.}~\bibnamefont {Riedl}}, \bibinfo {author} {\bibfnamefont {D.}~\bibnamefont {Amoroso}}, \bibinfo {author} {\bibfnamefont {S.}~\bibnamefont {Backes}}, \bibinfo {author} {\bibfnamefont {A.}~\bibnamefont {Razpopov}}, \bibinfo {author} {\bibfnamefont {T.~P.~T.}\ \bibnamefont {Nguyen}}, \bibinfo {author} {\bibfnamefont {K.}~\bibnamefont {Yamauchi}}, \bibinfo {author} {\bibfnamefont {P.}~\bibnamefont {Barone}}, \bibinfo {author} {\bibfnamefont {S.~M.}\ \bibnamefont {Winter}}, \bibinfo {author} {\bibfnamefont {S.}~\bibnamefont {Picozzi}},\ and\ \bibinfo {author} {\bibfnamefont {R.}~\bibnamefont {Valent\'{\i}}},\ }\bibfield  {title} {\bibinfo {title} {Microscopic origin of magnetism in monolayer $3d$ transition metal dihalides},\ }\href {https://doi.org/10.1103/PhysRevB.106.035156} {\bibfield  {journal} {\bibinfo  {journal} {Phys. Rev. B}\ }\textbf {\bibinfo {volume} {106}},\ \bibinfo {pages} {035156} (\bibinfo {year} {2022})}\BibitemShut {NoStop}%
\bibitem [{\citenamefont {Amini}\ \emph {et~al.}(2024)\citenamefont {Amini}, \citenamefont {Fumega}, \citenamefont {González-Herrero}, \citenamefont {Vaňo}, \citenamefont {Kezilebieke}, \citenamefont {Lado},\ and\ \citenamefont {Liljeroth}}]{Amini2024}%
  \BibitemOpen
  \bibfield  {author} {\bibinfo {author} {\bibfnamefont {M.}~\bibnamefont {Amini}}, \bibinfo {author} {\bibfnamefont {A.~O.}\ \bibnamefont {Fumega}}, \bibinfo {author} {\bibfnamefont {H.}~\bibnamefont {González-Herrero}}, \bibinfo {author} {\bibfnamefont {V.}~\bibnamefont {Vaňo}}, \bibinfo {author} {\bibfnamefont {S.}~\bibnamefont {Kezilebieke}}, \bibinfo {author} {\bibfnamefont {J.~L.}\ \bibnamefont {Lado}},\ and\ \bibinfo {author} {\bibfnamefont {P.}~\bibnamefont {Liljeroth}},\ }\bibfield  {title} {\bibinfo {title} {Atomic-scale visualization of multiferroicity in monolayer {N}i{I}$_2$},\ }\href {https://doi.org/https://doi.org/10.1002/adma.202311342} {\bibfield  {journal} {\bibinfo  {journal} {Adv. Mater.}\ }\textbf {\bibinfo {volume} {36}},\ \bibinfo {pages} {2311342} (\bibinfo {year} {2024})}\BibitemShut {NoStop}%
\bibitem [{\citenamefont {Posey}\ \emph {et~al.}(2024)\citenamefont {Posey}, \citenamefont {Turkel}, \citenamefont {Rezaee}, \citenamefont {Devarakonda}, \citenamefont {Kundu}, \citenamefont {Ong}, \citenamefont {Thinel}, \citenamefont {Chica}, \citenamefont {Vitalone}, \citenamefont {Jing}, \citenamefont {Xu}, \citenamefont {Needell}, \citenamefont {Meirzadeh}, \citenamefont {Feuer}, \citenamefont {Jindal}, \citenamefont {Cui}, \citenamefont {Valla}, \citenamefont {Thunstr\"{o}m}, \citenamefont {Yilmaz}, \citenamefont {Vescovo}, \citenamefont {Graf}, \citenamefont {Zhu}, \citenamefont {Scheie}, \citenamefont {May}, \citenamefont {Eriksson}, \citenamefont {Basov}, \citenamefont {Dean}, \citenamefont {Rubio}, \citenamefont {Kim}, \citenamefont {Ziebel}, \citenamefont {Millis}, \citenamefont {Pasupathy},\ and\ \citenamefont {Roy}}]{Posey2024}%
  \BibitemOpen
  \bibfield  {author} {\bibinfo {author} {\bibfnamefont {V.~A.}\ \bibnamefont {Posey}}, \bibinfo {author} {\bibfnamefont {S.}~\bibnamefont {Turkel}}, \bibinfo {author} {\bibfnamefont {M.}~\bibnamefont {Rezaee}}, \bibinfo {author} {\bibfnamefont {A.}~\bibnamefont {Devarakonda}}, \bibinfo {author} {\bibfnamefont {A.~K.}\ \bibnamefont {Kundu}}, \bibinfo {author} {\bibfnamefont {C.~S.}\ \bibnamefont {Ong}}, \bibinfo {author} {\bibfnamefont {M.}~\bibnamefont {Thinel}}, \bibinfo {author} {\bibfnamefont {D.~G.}\ \bibnamefont {Chica}}, \bibinfo {author} {\bibfnamefont {R.~A.}\ \bibnamefont {Vitalone}}, \bibinfo {author} {\bibfnamefont {R.}~\bibnamefont {Jing}}, \bibinfo {author} {\bibfnamefont {S.}~\bibnamefont {Xu}}, \bibinfo {author} {\bibfnamefont {D.~R.}\ \bibnamefont {Needell}}, \bibinfo {author} {\bibfnamefont {E.}~\bibnamefont {Meirzadeh}}, \bibinfo {author} {\bibfnamefont {M.~L.}\ \bibnamefont {Feuer}}, \bibinfo {author} {\bibfnamefont {A.}~\bibnamefont {Jindal}}, \bibinfo {author} {\bibfnamefont
  {X.}~\bibnamefont {Cui}}, \bibinfo {author} {\bibfnamefont {T.}~\bibnamefont {Valla}}, \bibinfo {author} {\bibfnamefont {P.}~\bibnamefont {Thunstr\"{o}m}}, \bibinfo {author} {\bibfnamefont {T.}~\bibnamefont {Yilmaz}}, \bibinfo {author} {\bibfnamefont {E.}~\bibnamefont {Vescovo}}, \bibinfo {author} {\bibfnamefont {D.}~\bibnamefont {Graf}}, \bibinfo {author} {\bibfnamefont {X.}~\bibnamefont {Zhu}}, \bibinfo {author} {\bibfnamefont {A.}~\bibnamefont {Scheie}}, \bibinfo {author} {\bibfnamefont {A.~F.}\ \bibnamefont {May}}, \bibinfo {author} {\bibfnamefont {O.}~\bibnamefont {Eriksson}}, \bibinfo {author} {\bibfnamefont {D.~N.}\ \bibnamefont {Basov}}, \bibinfo {author} {\bibfnamefont {C.~R.}\ \bibnamefont {Dean}}, \bibinfo {author} {\bibfnamefont {A.}~\bibnamefont {Rubio}}, \bibinfo {author} {\bibfnamefont {P.}~\bibnamefont {Kim}}, \bibinfo {author} {\bibfnamefont {M.~E.}\ \bibnamefont {Ziebel}}, \bibinfo {author} {\bibfnamefont {A.~J.}\ \bibnamefont {Millis}}, \bibinfo {author} {\bibfnamefont {A.~N.}\
  \bibnamefont {Pasupathy}},\ and\ \bibinfo {author} {\bibfnamefont {X.}~\bibnamefont {Roy}},\ }\bibfield  {title} {\bibinfo {title} {Two-dimensional heavy fermions in the van der {W}aals metal {C}e{S}i{I}},\ }\href {https://doi.org/10.1038/s41586-023-06868-x} {\bibfield  {journal} {\bibinfo  {journal} {Nature}\ }\textbf {\bibinfo {volume} {625}},\ \bibinfo {pages} {483–488} (\bibinfo {year} {2024})}\BibitemShut {NoStop}%
\bibitem [{\citenamefont {Fumega}\ and\ \citenamefont {Lado}(2024)}]{Fumega2024}%
  \BibitemOpen
  \bibfield  {author} {\bibinfo {author} {\bibfnamefont {A.~O.}\ \bibnamefont {Fumega}}\ and\ \bibinfo {author} {\bibfnamefont {J.~L.}\ \bibnamefont {Lado}},\ }\bibfield  {title} {\bibinfo {title} {Nature of the unconventional heavy-fermion {K}ondo state in monolayer {C}e{S}i{I}},\ }\href {https://doi.org/10.1021/acs.nanolett.4c00619} {\bibfield  {journal} {\bibinfo  {journal} {Nano Lett.}\ }\textbf {\bibinfo {volume} {24}},\ \bibinfo {pages} {4272–4278} (\bibinfo {year} {2024})}\BibitemShut {NoStop}%
\bibitem [{\citenamefont {Vijayvargia}\ and\ \citenamefont {Erten}(2024)}]{PhysRevB.109.L201118}%
  \BibitemOpen
  \bibfield  {author} {\bibinfo {author} {\bibfnamefont {A.}~\bibnamefont {Vijayvargia}}\ and\ \bibinfo {author} {\bibfnamefont {O.}~\bibnamefont {Erten}},\ }\bibfield  {title} {\bibinfo {title} {Nematic heavy fermions and coexisting magnetic order in {C}e{S}i{I}},\ }\href {https://doi.org/10.1103/PhysRevB.109.L201118} {\bibfield  {journal} {\bibinfo  {journal} {Phys. Rev. B}\ }\textbf {\bibinfo {volume} {109}},\ \bibinfo {pages} {L201118} (\bibinfo {year} {2024})}\BibitemShut {NoStop}%
\bibitem [{\citenamefont {Vaňo}\ \emph {et~al.}(2021)\citenamefont {Vaňo}, \citenamefont {Amini}, \citenamefont {Ganguli}, \citenamefont {Chen}, \citenamefont {Lado}, \citenamefont {Kezilebieke},\ and\ \citenamefont {Liljeroth}}]{Vano2021}%
  \BibitemOpen
  \bibfield  {author} {\bibinfo {author} {\bibfnamefont {V.}~\bibnamefont {Vaňo}}, \bibinfo {author} {\bibfnamefont {M.}~\bibnamefont {Amini}}, \bibinfo {author} {\bibfnamefont {S.~C.}\ \bibnamefont {Ganguli}}, \bibinfo {author} {\bibfnamefont {G.}~\bibnamefont {Chen}}, \bibinfo {author} {\bibfnamefont {J.~L.}\ \bibnamefont {Lado}}, \bibinfo {author} {\bibfnamefont {S.}~\bibnamefont {Kezilebieke}},\ and\ \bibinfo {author} {\bibfnamefont {P.}~\bibnamefont {Liljeroth}},\ }\bibfield  {title} {\bibinfo {title} {Artificial heavy fermions in a van der {W}aals heterostructure},\ }\href {https://doi.org/10.1038/s41586-021-04021-0} {\bibfield  {journal} {\bibinfo  {journal} {Nature}\ }\textbf {\bibinfo {volume} {599}},\ \bibinfo {pages} {582–586} (\bibinfo {year} {2021})}\BibitemShut {NoStop}%
\bibitem [{\citenamefont {Wan}\ \emph {et~al.}(2023)\citenamefont {Wan}, \citenamefont {Harsh}, \citenamefont {Meninno}, \citenamefont {Dreher}, \citenamefont {Sajan}, \citenamefont {Guo}, \citenamefont {Errea}, \citenamefont {de~Juan},\ and\ \citenamefont {Ugeda}}]{Wan2023}%
  \BibitemOpen
  \bibfield  {author} {\bibinfo {author} {\bibfnamefont {W.}~\bibnamefont {Wan}}, \bibinfo {author} {\bibfnamefont {R.}~\bibnamefont {Harsh}}, \bibinfo {author} {\bibfnamefont {A.}~\bibnamefont {Meninno}}, \bibinfo {author} {\bibfnamefont {P.}~\bibnamefont {Dreher}}, \bibinfo {author} {\bibfnamefont {S.}~\bibnamefont {Sajan}}, \bibinfo {author} {\bibfnamefont {H.}~\bibnamefont {Guo}}, \bibinfo {author} {\bibfnamefont {I.}~\bibnamefont {Errea}}, \bibinfo {author} {\bibfnamefont {F.}~\bibnamefont {de~Juan}},\ and\ \bibinfo {author} {\bibfnamefont {M.~M.}\ \bibnamefont {Ugeda}},\ }\bibfield  {title} {\bibinfo {title} {Evidence for ground state coherence in a two-dimensional {K}ondo lattice},\ }\href {https://doi.org/10.1038/s41467-023-42803-4} {\bibfield  {journal} {\bibinfo  {journal} {Nat. Commun.}\ }\textbf {\bibinfo {volume} {14}},\ \bibinfo {pages} {7005} (\bibinfo {year} {2023})}\BibitemShut {NoStop}%
\bibitem [{\citenamefont {Ma{\~n}as-Valero}\ \emph {et~al.}(2021)\citenamefont {Ma{\~n}as-Valero}, \citenamefont {Huddart}, \citenamefont {Lancaster}, \citenamefont {Coronado},\ and\ \citenamefont {Pratt}}]{MaasValero2021}%
  \BibitemOpen
  \bibfield  {author} {\bibinfo {author} {\bibfnamefont {S.}~\bibnamefont {Ma{\~n}as-Valero}}, \bibinfo {author} {\bibfnamefont {B.~M.}\ \bibnamefont {Huddart}}, \bibinfo {author} {\bibfnamefont {T.}~\bibnamefont {Lancaster}}, \bibinfo {author} {\bibfnamefont {E.}~\bibnamefont {Coronado}},\ and\ \bibinfo {author} {\bibfnamefont {F.~L.}\ \bibnamefont {Pratt}},\ }\bibfield  {title} {\bibinfo {title} {Quantum phases and spin liquid properties of 1{T}-{T}a{S}$_2$},\ }\href {https://doi.org/10.1038/s41535-021-00367-w} {\bibfield  {journal} {\bibinfo  {journal} {npj Quantum Materials}\ }\textbf {\bibinfo {volume} {6}},\ \bibinfo {pages} {69} (\bibinfo {year} {2021})}\BibitemShut {NoStop}%
\bibitem [{\citenamefont {Wang}\ \emph {et~al.}(2024{\natexlab{a}})\citenamefont {Wang}, \citenamefont {Li}, \citenamefont {Luo}, \citenamefont {Gao}, \citenamefont {Han}, \citenamefont {Jiang}, \citenamefont {Tang}, \citenamefont {Ju}, \citenamefont {Li}, \citenamefont {Lv}, \citenamefont {Cui}, \citenamefont {Yang}, \citenamefont {Sun}, \citenamefont {Zhu}, \citenamefont {Gao}, \citenamefont {Lu}, \citenamefont {Sun}, \citenamefont {Xu}, \citenamefont {Xiong},\ and\ \citenamefont {Cao}}]{Wang2024}%
  \BibitemOpen
  \bibfield  {author} {\bibinfo {author} {\bibfnamefont {Y.}~\bibnamefont {Wang}}, \bibinfo {author} {\bibfnamefont {Z.}~\bibnamefont {Li}}, \bibinfo {author} {\bibfnamefont {X.}~\bibnamefont {Luo}}, \bibinfo {author} {\bibfnamefont {J.}~\bibnamefont {Gao}}, \bibinfo {author} {\bibfnamefont {Y.}~\bibnamefont {Han}}, \bibinfo {author} {\bibfnamefont {J.}~\bibnamefont {Jiang}}, \bibinfo {author} {\bibfnamefont {J.}~\bibnamefont {Tang}}, \bibinfo {author} {\bibfnamefont {H.}~\bibnamefont {Ju}}, \bibinfo {author} {\bibfnamefont {T.}~\bibnamefont {Li}}, \bibinfo {author} {\bibfnamefont {R.}~\bibnamefont {Lv}}, \bibinfo {author} {\bibfnamefont {S.}~\bibnamefont {Cui}}, \bibinfo {author} {\bibfnamefont {Y.}~\bibnamefont {Yang}}, \bibinfo {author} {\bibfnamefont {Y.}~\bibnamefont {Sun}}, \bibinfo {author} {\bibfnamefont {J.}~\bibnamefont {Zhu}}, \bibinfo {author} {\bibfnamefont {X.}~\bibnamefont {Gao}}, \bibinfo {author} {\bibfnamefont {W.}~\bibnamefont {Lu}}, \bibinfo {author} {\bibfnamefont {Z.}~\bibnamefont
  {Sun}}, \bibinfo {author} {\bibfnamefont {H.}~\bibnamefont {Xu}}, \bibinfo {author} {\bibfnamefont {Y.}~\bibnamefont {Xiong}},\ and\ \bibinfo {author} {\bibfnamefont {L.}~\bibnamefont {Cao}},\ }\bibfield  {title} {\bibinfo {title} {Dualistic insulator states in 1{T}-{T}a{S}$_2$ crystals},\ }\href {https://doi.org/10.1038/s41467-024-47728-0} {\bibfield  {journal} {\bibinfo  {journal} {Nat. Commun.}\ }\textbf {\bibinfo {volume} {15}},\ \bibinfo {pages} {3425} (\bibinfo {year} {2024}{\natexlab{a}})}\BibitemShut {NoStop}%
\bibitem [{\citenamefont {Law}\ and\ \citenamefont {Lee}(2017)}]{Law2017}%
  \BibitemOpen
  \bibfield  {author} {\bibinfo {author} {\bibfnamefont {K.~T.}\ \bibnamefont {Law}}\ and\ \bibinfo {author} {\bibfnamefont {P.~A.}\ \bibnamefont {Lee}},\ }\bibfield  {title} {\bibinfo {title} {1{T}-{T}a{S}$_2$ as a quantum spin liquid},\ }\href {https://doi.org/10.1073/pnas.1706769114} {\bibfield  {journal} {\bibinfo  {journal} {Proc. Natl. Acad. Sci. U.S.A.}\ }\textbf {\bibinfo {volume} {114}},\ \bibinfo {pages} {6996–7000} (\bibinfo {year} {2017})}\BibitemShut {NoStop}%
\bibitem [{\citenamefont {Chen}\ \emph {et~al.}(2022{\natexlab{a}})\citenamefont {Chen}, \citenamefont {He}, \citenamefont {Ruan}, \citenamefont {Hwang}, \citenamefont {Tang}, \citenamefont {Lee}, \citenamefont {Wu}, \citenamefont {Zhu}, \citenamefont {Zhang}, \citenamefont {Ryu}, \citenamefont {Wang}, \citenamefont {Louie}, \citenamefont {Shen}, \citenamefont {Mo}, \citenamefont {Lee},\ and\ \citenamefont {Crommie}}]{Chen2022}%
  \BibitemOpen
  \bibfield  {author} {\bibinfo {author} {\bibfnamefont {Y.}~\bibnamefont {Chen}}, \bibinfo {author} {\bibfnamefont {W.-Y.}\ \bibnamefont {He}}, \bibinfo {author} {\bibfnamefont {W.}~\bibnamefont {Ruan}}, \bibinfo {author} {\bibfnamefont {J.}~\bibnamefont {Hwang}}, \bibinfo {author} {\bibfnamefont {S.}~\bibnamefont {Tang}}, \bibinfo {author} {\bibfnamefont {R.~L.}\ \bibnamefont {Lee}}, \bibinfo {author} {\bibfnamefont {M.}~\bibnamefont {Wu}}, \bibinfo {author} {\bibfnamefont {T.}~\bibnamefont {Zhu}}, \bibinfo {author} {\bibfnamefont {C.}~\bibnamefont {Zhang}}, \bibinfo {author} {\bibfnamefont {H.}~\bibnamefont {Ryu}}, \bibinfo {author} {\bibfnamefont {F.}~\bibnamefont {Wang}}, \bibinfo {author} {\bibfnamefont {S.~G.}\ \bibnamefont {Louie}}, \bibinfo {author} {\bibfnamefont {Z.-X.}\ \bibnamefont {Shen}}, \bibinfo {author} {\bibfnamefont {S.-K.}\ \bibnamefont {Mo}}, \bibinfo {author} {\bibfnamefont {P.~A.}\ \bibnamefont {Lee}},\ and\ \bibinfo {author} {\bibfnamefont {M.~F.}\ \bibnamefont {Crommie}},\ }\bibfield
   {title} {\bibinfo {title} {Evidence for a spinon {K}ondo effect in cobalt atoms on single-layer 1{T}-{T}a{S}e$_2$},\ }\href {https://doi.org/10.1038/s41567-022-01751-4} {\bibfield  {journal} {\bibinfo  {journal} {Nat. Phys.}\ }\textbf {\bibinfo {volume} {18}},\ \bibinfo {pages} {1335–1340} (\bibinfo {year} {2022}{\natexlab{a}})}\BibitemShut {NoStop}%
\bibitem [{\citenamefont {Ruan}\ \emph {et~al.}(2021)\citenamefont {Ruan}, \citenamefont {Chen}, \citenamefont {Tang}, \citenamefont {Hwang}, \citenamefont {Tsai}, \citenamefont {Lee}, \citenamefont {Wu}, \citenamefont {Ryu}, \citenamefont {Kahn}, \citenamefont {Liou}, \citenamefont {Jia}, \citenamefont {Aikawa}, \citenamefont {Hwang}, \citenamefont {Wang}, \citenamefont {Choi}, \citenamefont {Louie}, \citenamefont {Lee}, \citenamefont {Shen}, \citenamefont {Mo},\ and\ \citenamefont {Crommie}}]{Ruan2021}%
  \BibitemOpen
  \bibfield  {author} {\bibinfo {author} {\bibfnamefont {W.}~\bibnamefont {Ruan}}, \bibinfo {author} {\bibfnamefont {Y.}~\bibnamefont {Chen}}, \bibinfo {author} {\bibfnamefont {S.}~\bibnamefont {Tang}}, \bibinfo {author} {\bibfnamefont {J.}~\bibnamefont {Hwang}}, \bibinfo {author} {\bibfnamefont {H.-Z.}\ \bibnamefont {Tsai}}, \bibinfo {author} {\bibfnamefont {R.~L.}\ \bibnamefont {Lee}}, \bibinfo {author} {\bibfnamefont {M.}~\bibnamefont {Wu}}, \bibinfo {author} {\bibfnamefont {H.}~\bibnamefont {Ryu}}, \bibinfo {author} {\bibfnamefont {S.}~\bibnamefont {Kahn}}, \bibinfo {author} {\bibfnamefont {F.}~\bibnamefont {Liou}}, \bibinfo {author} {\bibfnamefont {C.}~\bibnamefont {Jia}}, \bibinfo {author} {\bibfnamefont {A.}~\bibnamefont {Aikawa}}, \bibinfo {author} {\bibfnamefont {C.}~\bibnamefont {Hwang}}, \bibinfo {author} {\bibfnamefont {F.}~\bibnamefont {Wang}}, \bibinfo {author} {\bibfnamefont {Y.}~\bibnamefont {Choi}}, \bibinfo {author} {\bibfnamefont {S.~G.}\ \bibnamefont {Louie}}, \bibinfo {author}
  {\bibfnamefont {P.~A.}\ \bibnamefont {Lee}}, \bibinfo {author} {\bibfnamefont {Z.-X.}\ \bibnamefont {Shen}}, \bibinfo {author} {\bibfnamefont {S.-K.}\ \bibnamefont {Mo}},\ and\ \bibinfo {author} {\bibfnamefont {M.~F.}\ \bibnamefont {Crommie}},\ }\bibfield  {title} {\bibinfo {title} {Evidence for quantum spin liquid behaviour in single-layer 1{T}-{T}a{S}e$_2$ from scanning tunnelling microscopy},\ }\href {https://doi.org/10.1038/s41567-021-01321-0} {\bibfield  {journal} {\bibinfo  {journal} {Nat. Phys.}\ }\textbf {\bibinfo {volume} {17}},\ \bibinfo {pages} {1154–1161} (\bibinfo {year} {2021})}\BibitemShut {NoStop}%
\bibitem [{\citenamefont {Bernevig}\ \emph {et~al.}(2006)\citenamefont {Bernevig}, \citenamefont {Hughes},\ and\ \citenamefont {Zhang}}]{BHZ2006}%
  \BibitemOpen
  \bibfield  {author} {\bibinfo {author} {\bibfnamefont {B.~A.}\ \bibnamefont {Bernevig}}, \bibinfo {author} {\bibfnamefont {T.~L.}\ \bibnamefont {Hughes}},\ and\ \bibinfo {author} {\bibfnamefont {S.-C.}\ \bibnamefont {Zhang}},\ }\bibfield  {title} {\bibinfo {title} {Quantum spin {H}all effect and topological phase transition in {H}g{T}e quantum wells},\ }\href {https://doi.org/10.1126/science.1133734} {\bibfield  {journal} {\bibinfo  {journal} {Science}\ }\textbf {\bibinfo {volume} {314}},\ \bibinfo {pages} {1757} (\bibinfo {year} {2006})}\BibitemShut {NoStop}%
\bibitem [{\citenamefont {König}\ \emph {et~al.}(2007)\citenamefont {König}, \citenamefont {Wiedmann}, \citenamefont {Brune}, \citenamefont {Roth}, \citenamefont {Buhmann}, \citenamefont {Molenkamp}, \citenamefont {Qi},\ and\ \citenamefont {Zhang}}]{konig2007}%
  \BibitemOpen
  \bibfield  {author} {\bibinfo {author} {\bibfnamefont {M.}~\bibnamefont {König}}, \bibinfo {author} {\bibfnamefont {S.}~\bibnamefont {Wiedmann}}, \bibinfo {author} {\bibfnamefont {C.}~\bibnamefont {Brune}}, \bibinfo {author} {\bibfnamefont {A.}~\bibnamefont {Roth}}, \bibinfo {author} {\bibfnamefont {H.}~\bibnamefont {Buhmann}}, \bibinfo {author} {\bibfnamefont {L.~W.}\ \bibnamefont {Molenkamp}}, \bibinfo {author} {\bibfnamefont {X.-L.}\ \bibnamefont {Qi}},\ and\ \bibinfo {author} {\bibfnamefont {S.-C.}\ \bibnamefont {Zhang}},\ }\bibfield  {title} {\bibinfo {title} {Quantum spin {H}all insulator state in {H}g{T}e quantum wells},\ }\href {https://doi.org/10.1126/science.1148047} {\bibfield  {journal} {\bibinfo  {journal} {Science}\ }\textbf {\bibinfo {volume} {318}},\ \bibinfo {pages} {766} (\bibinfo {year} {2007})}\BibitemShut {NoStop}%
\bibitem [{\citenamefont {Xia}\ \emph {et~al.}(2009)\citenamefont {Xia}, \citenamefont {Qian}, \citenamefont {Hsieh}, \citenamefont {Wray}, \citenamefont {Pal}, \citenamefont {Lin}, \citenamefont {Bansil}, \citenamefont {Grauer}, \citenamefont {Hor}, \citenamefont {Cava},\ and\ \citenamefont {Hasan}}]{xia2009}%
  \BibitemOpen
  \bibfield  {author} {\bibinfo {author} {\bibfnamefont {Y.}~\bibnamefont {Xia}}, \bibinfo {author} {\bibfnamefont {D.}~\bibnamefont {Qian}}, \bibinfo {author} {\bibfnamefont {D.}~\bibnamefont {Hsieh}}, \bibinfo {author} {\bibfnamefont {L.}~\bibnamefont {Wray}}, \bibinfo {author} {\bibfnamefont {A.}~\bibnamefont {Pal}}, \bibinfo {author} {\bibfnamefont {H.}~\bibnamefont {Lin}}, \bibinfo {author} {\bibfnamefont {A.}~\bibnamefont {Bansil}}, \bibinfo {author} {\bibfnamefont {D.}~\bibnamefont {Grauer}}, \bibinfo {author} {\bibfnamefont {Y.~S.}\ \bibnamefont {Hor}}, \bibinfo {author} {\bibfnamefont {R.~J.}\ \bibnamefont {Cava}},\ and\ \bibinfo {author} {\bibfnamefont {M.~Z.}\ \bibnamefont {Hasan}},\ }\bibfield  {title} {\bibinfo {title} {Observation of a large-gap topological-insulator class with a single {D}irac cone on the surface},\ }\href {https://doi.org/10.1038/nphys1274} {\bibfield  {journal} {\bibinfo  {journal} {Nat. Phys.}\ }\textbf {\bibinfo {volume} {5}},\ \bibinfo {pages} {398} (\bibinfo {year}
  {2009})}\BibitemShut {NoStop}%
\bibitem [{\citenamefont {Fei}\ \emph {et~al.}(2017)\citenamefont {Fei}, \citenamefont {Palomaki}, \citenamefont {Wu}, \citenamefont {Zhao}, \citenamefont {Cai}, \citenamefont {Sun}, \citenamefont {Nguyen}, \citenamefont {Finney}, \citenamefont {Xu},\ and\ \citenamefont {Cobden}}]{fei2017}%
  \BibitemOpen
  \bibfield  {author} {\bibinfo {author} {\bibfnamefont {Z.}~\bibnamefont {Fei}}, \bibinfo {author} {\bibfnamefont {T.}~\bibnamefont {Palomaki}}, \bibinfo {author} {\bibfnamefont {S.}~\bibnamefont {Wu}}, \bibinfo {author} {\bibfnamefont {W.}~\bibnamefont {Zhao}}, \bibinfo {author} {\bibfnamefont {X.}~\bibnamefont {Cai}}, \bibinfo {author} {\bibfnamefont {B.}~\bibnamefont {Sun}}, \bibinfo {author} {\bibfnamefont {P.}~\bibnamefont {Nguyen}}, \bibinfo {author} {\bibfnamefont {J.}~\bibnamefont {Finney}}, \bibinfo {author} {\bibfnamefont {X.}~\bibnamefont {Xu}},\ and\ \bibinfo {author} {\bibfnamefont {D.~H.}\ \bibnamefont {Cobden}},\ }\bibfield  {title} {\bibinfo {title} {Edge conduction in monolayer {WT}e$_2$},\ }\href {https://doi.org/10.1038/nphys4091} {\bibfield  {journal} {\bibinfo  {journal} {Nat. Phys.}\ }\textbf {\bibinfo {volume} {13}},\ \bibinfo {pages} {677} (\bibinfo {year} {2017})}\BibitemShut {NoStop}%
\bibitem [{\citenamefont {Fu}\ and\ \citenamefont {Kane}(2008)}]{fu2008}%
  \BibitemOpen
  \bibfield  {author} {\bibinfo {author} {\bibfnamefont {L.}~\bibnamefont {Fu}}\ and\ \bibinfo {author} {\bibfnamefont {C.~L.}\ \bibnamefont {Kane}},\ }\bibfield  {title} {\bibinfo {title} {Superconducting proximity effect and {M}ajorana fermions at the surface of a topological insulator},\ }\href {https://doi.org/10.1103/PhysRevLett.100.096407} {\bibfield  {journal} {\bibinfo  {journal} {Phys. Rev. Lett.}\ }\textbf {\bibinfo {volume} {100}},\ \bibinfo {pages} {096407} (\bibinfo {year} {2008})}\BibitemShut {NoStop}%
\bibitem [{\citenamefont {He}\ \emph {et~al.}(2022)\citenamefont {He}, \citenamefont {Hughes}, \citenamefont {Armitage}, \citenamefont {Tokura},\ and\ \citenamefont {Wang}}]{he2022}%
  \BibitemOpen
  \bibfield  {author} {\bibinfo {author} {\bibfnamefont {Q.~L.}\ \bibnamefont {He}}, \bibinfo {author} {\bibfnamefont {T.~L.}\ \bibnamefont {Hughes}}, \bibinfo {author} {\bibfnamefont {N.~P.}\ \bibnamefont {Armitage}}, \bibinfo {author} {\bibfnamefont {Y.}~\bibnamefont {Tokura}},\ and\ \bibinfo {author} {\bibfnamefont {K.~L.}\ \bibnamefont {Wang}},\ }\bibfield  {title} {\bibinfo {title} {Topological spintronics and magnetoelectronics},\ }\href {https://doi.org/10.1038/s41563-021-01138-5} {\bibfield  {journal} {\bibinfo  {journal} {Nat. Mater.}\ }\textbf {\bibinfo {volume} {21}},\ \bibinfo {pages} {15} (\bibinfo {year} {2022})}\BibitemShut {NoStop}%
\bibitem [{\citenamefont {Tang}\ \emph {et~al.}(2017)\citenamefont {Tang}, \citenamefont {Zhang}, \citenamefont {Wong}, \citenamefont {Pedramrazi}, \citenamefont {Tsai}, \citenamefont {Jia}, \citenamefont {Moritz}, \citenamefont {Claassen}, \citenamefont {Ryu}, \citenamefont {Kahn}, \citenamefont {Jiang}, \citenamefont {Yan}, \citenamefont {Hashimoto}, \citenamefont {Lu}, \citenamefont {Moore}, \citenamefont {Hwang}, \citenamefont {Hwang}, \citenamefont {Hussain}, \citenamefont {Chen}, \citenamefont {Ugeda}, \citenamefont {Liu}, \citenamefont {Xie}, \citenamefont {Devereaux}, \citenamefont {Crommie}, \citenamefont {Mo},\ and\ \citenamefont {Shen}}]{Tang2017}%
  \BibitemOpen
  \bibfield  {author} {\bibinfo {author} {\bibfnamefont {S.}~\bibnamefont {Tang}}, \bibinfo {author} {\bibfnamefont {C.}~\bibnamefont {Zhang}}, \bibinfo {author} {\bibfnamefont {D.}~\bibnamefont {Wong}}, \bibinfo {author} {\bibfnamefont {Z.}~\bibnamefont {Pedramrazi}}, \bibinfo {author} {\bibfnamefont {H.-Z.}\ \bibnamefont {Tsai}}, \bibinfo {author} {\bibfnamefont {C.}~\bibnamefont {Jia}}, \bibinfo {author} {\bibfnamefont {B.}~\bibnamefont {Moritz}}, \bibinfo {author} {\bibfnamefont {M.}~\bibnamefont {Claassen}}, \bibinfo {author} {\bibfnamefont {H.}~\bibnamefont {Ryu}}, \bibinfo {author} {\bibfnamefont {S.}~\bibnamefont {Kahn}}, \bibinfo {author} {\bibfnamefont {J.}~\bibnamefont {Jiang}}, \bibinfo {author} {\bibfnamefont {H.}~\bibnamefont {Yan}}, \bibinfo {author} {\bibfnamefont {M.}~\bibnamefont {Hashimoto}}, \bibinfo {author} {\bibfnamefont {D.}~\bibnamefont {Lu}}, \bibinfo {author} {\bibfnamefont {R.~G.}\ \bibnamefont {Moore}}, \bibinfo {author} {\bibfnamefont {C.-C.}\ \bibnamefont {Hwang}}, \bibinfo
  {author} {\bibfnamefont {C.}~\bibnamefont {Hwang}}, \bibinfo {author} {\bibfnamefont {Z.}~\bibnamefont {Hussain}}, \bibinfo {author} {\bibfnamefont {Y.}~\bibnamefont {Chen}}, \bibinfo {author} {\bibfnamefont {M.~M.}\ \bibnamefont {Ugeda}}, \bibinfo {author} {\bibfnamefont {Z.}~\bibnamefont {Liu}}, \bibinfo {author} {\bibfnamefont {X.}~\bibnamefont {Xie}}, \bibinfo {author} {\bibfnamefont {T.~P.}\ \bibnamefont {Devereaux}}, \bibinfo {author} {\bibfnamefont {M.~F.}\ \bibnamefont {Crommie}}, \bibinfo {author} {\bibfnamefont {S.-K.}\ \bibnamefont {Mo}},\ and\ \bibinfo {author} {\bibfnamefont {Z.-X.}\ \bibnamefont {Shen}},\ }\bibfield  {title} {\bibinfo {title} {Quantum spin {H}all state in monolayer 1{T}’-{WT}e2},\ }\href {https://doi.org/10.1038/nphys4174} {\bibfield  {journal} {\bibinfo  {journal} {Nat. Phys.}\ }\textbf {\bibinfo {volume} {13}},\ \bibinfo {pages} {683–687} (\bibinfo {year} {2017})}\BibitemShut {NoStop}%
\bibitem [{\citenamefont {Wu}\ \emph {et~al.}(2018)\citenamefont {Wu}, \citenamefont {Fatemi}, \citenamefont {Gibson}, \citenamefont {Watanabe}, \citenamefont {Taniguchi}, \citenamefont {Cava},\ and\ \citenamefont {Jarillo-Herrero}}]{Wu2018}%
  \BibitemOpen
  \bibfield  {author} {\bibinfo {author} {\bibfnamefont {S.}~\bibnamefont {Wu}}, \bibinfo {author} {\bibfnamefont {V.}~\bibnamefont {Fatemi}}, \bibinfo {author} {\bibfnamefont {Q.~D.}\ \bibnamefont {Gibson}}, \bibinfo {author} {\bibfnamefont {K.}~\bibnamefont {Watanabe}}, \bibinfo {author} {\bibfnamefont {T.}~\bibnamefont {Taniguchi}}, \bibinfo {author} {\bibfnamefont {R.~J.}\ \bibnamefont {Cava}},\ and\ \bibinfo {author} {\bibfnamefont {P.}~\bibnamefont {Jarillo-Herrero}},\ }\bibfield  {title} {\bibinfo {title} {Observation of the quantum spin {H}all effect up to 100 kelvin in a monolayer crystal},\ }\href {https://doi.org/10.1126/science.aan6003} {\bibfield  {journal} {\bibinfo  {journal} {Science}\ }\textbf {\bibinfo {volume} {359}},\ \bibinfo {pages} {76–79} (\bibinfo {year} {2018})}\BibitemShut {NoStop}%
\bibitem [{\citenamefont {Ueda}\ \emph {et~al.}(2013)\citenamefont {Ueda}, \citenamefont {Kawakami},\ and\ \citenamefont {Sigrist}}]{Ueda2013}%
  \BibitemOpen
  \bibfield  {author} {\bibinfo {author} {\bibfnamefont {S.}~\bibnamefont {Ueda}}, \bibinfo {author} {\bibfnamefont {N.}~\bibnamefont {Kawakami}},\ and\ \bibinfo {author} {\bibfnamefont {M.}~\bibnamefont {Sigrist}},\ }\bibfield  {title} {\bibinfo {title} {Proximity effects in a topological-insulator/{M}ott-insulator heterostructure},\ }\href {https://doi.org/10.1103/PhysRevB.87.161108} {\bibfield  {journal} {\bibinfo  {journal} {Phys. Rev. B}\ }\textbf {\bibinfo {volume} {87}},\ \bibinfo {pages} {161108} (\bibinfo {year} {2013})}\BibitemShut {NoStop}%
\bibitem [{\citenamefont {Yokoyama}\ \emph {et~al.}(2010)\citenamefont {Yokoyama}, \citenamefont {Tanaka},\ and\ \citenamefont {Nagaosa}}]{Yokoyama2010}%
  \BibitemOpen
  \bibfield  {author} {\bibinfo {author} {\bibfnamefont {T.}~\bibnamefont {Yokoyama}}, \bibinfo {author} {\bibfnamefont {Y.}~\bibnamefont {Tanaka}},\ and\ \bibinfo {author} {\bibfnamefont {N.}~\bibnamefont {Nagaosa}},\ }\bibfield  {title} {\bibinfo {title} {Anomalous magnetoresistance of a two-dimensional ferromagnet/ferromagnet junction on the surface of a topological insulator},\ }\href {https://doi.org/10.1103/PhysRevB.81.121401} {\bibfield  {journal} {\bibinfo  {journal} {Phys. Rev. B}\ }\textbf {\bibinfo {volume} {81}},\ \bibinfo {pages} {121401} (\bibinfo {year} {2010})}\BibitemShut {NoStop}%
\bibitem [{\citenamefont {Li}\ \emph {et~al.}(2013)\citenamefont {Li}, \citenamefont {Zhang}, \citenamefont {Wu}, \citenamefont {Chen}, \citenamefont {Culcer},\ and\ \citenamefont {Zhang}}]{li2013}%
  \BibitemOpen
  \bibfield  {author} {\bibinfo {author} {\bibfnamefont {X.-G.}\ \bibnamefont {Li}}, \bibinfo {author} {\bibfnamefont {G.-F.}\ \bibnamefont {Zhang}}, \bibinfo {author} {\bibfnamefont {G.-F.}\ \bibnamefont {Wu}}, \bibinfo {author} {\bibfnamefont {H.}~\bibnamefont {Chen}}, \bibinfo {author} {\bibfnamefont {D.}~\bibnamefont {Culcer}},\ and\ \bibinfo {author} {\bibfnamefont {Z.-Y.}\ \bibnamefont {Zhang}},\ }\bibfield  {title} {\bibinfo {title} {Proximity effects in topological insulator heterostructures},\ }\href {https://doi.org/10.1088/1674-1056/22/9/097306} {\bibfield  {journal} {\bibinfo  {journal} {Chin. Phys. B}\ }\textbf {\bibinfo {volume} {22}},\ \bibinfo {pages} {097306} (\bibinfo {year} {2013})}\BibitemShut {NoStop}%
\bibitem [{\citenamefont {Teixeira}\ \emph {et~al.}(2019)\citenamefont {Teixeira}, \citenamefont {Kuzmanovski}, \citenamefont {Black-Schaffer},\ and\ \citenamefont {Dias~da Silva}}]{Teixeira2019}%
  \BibitemOpen
  \bibfield  {author} {\bibinfo {author} {\bibfnamefont {R.~L. R.~C.}\ \bibnamefont {Teixeira}}, \bibinfo {author} {\bibfnamefont {D.}~\bibnamefont {Kuzmanovski}}, \bibinfo {author} {\bibfnamefont {A.~M.}\ \bibnamefont {Black-Schaffer}},\ and\ \bibinfo {author} {\bibfnamefont {L.~G. G.~V.}\ \bibnamefont {Dias~da Silva}},\ }\bibfield  {title} {\bibinfo {title} {Gap oscillations and {M}ajorana bound states in magnetic chains on superconducting honeycomb lattices},\ }\href {https://doi.org/10.1103/PhysRevB.99.035127} {\bibfield  {journal} {\bibinfo  {journal} {Phys. Rev. B}\ }\textbf {\bibinfo {volume} {99}},\ \bibinfo {pages} {035127} (\bibinfo {year} {2019})}\BibitemShut {NoStop}%
\bibitem [{\citenamefont {Li}\ \emph {et~al.}(2015)\citenamefont {Li}, \citenamefont {Chan},\ and\ \citenamefont {Yao}}]{Li2015}%
  \BibitemOpen
  \bibfield  {author} {\bibinfo {author} {\bibfnamefont {Z.-X.}\ \bibnamefont {Li}}, \bibinfo {author} {\bibfnamefont {C.}~\bibnamefont {Chan}},\ and\ \bibinfo {author} {\bibfnamefont {H.}~\bibnamefont {Yao}},\ }\bibfield  {title} {\bibinfo {title} {Realizing {M}ajorana zero modes by proximity effect between topological insulators and $d$-wave high-temperature superconductors},\ }\href {https://doi.org/10.1103/PhysRevB.91.235143} {\bibfield  {journal} {\bibinfo  {journal} {Phys. Rev. B}\ }\textbf {\bibinfo {volume} {91}},\ \bibinfo {pages} {235143} (\bibinfo {year} {2015})}\BibitemShut {NoStop}%
\bibitem [{\citenamefont {Stagraczy\ifmmode~\acute{n}\else \'{n}\fi{}ski}\ \emph {et~al.}(2016)\citenamefont {Stagraczy\ifmmode~\acute{n}\else \'{n}\fi{}ski}, \citenamefont {Chotorlishvili}, \citenamefont {Dugaev}, \citenamefont {Jia}, \citenamefont {Ernst}, \citenamefont {Komnik},\ and\ \citenamefont {Berakdar}}]{Stagraczynski2016}%
  \BibitemOpen
  \bibfield  {author} {\bibinfo {author} {\bibfnamefont {S.}~\bibnamefont {Stagraczy\ifmmode~\acute{n}\else \'{n}\fi{}ski}}, \bibinfo {author} {\bibfnamefont {L.}~\bibnamefont {Chotorlishvili}}, \bibinfo {author} {\bibfnamefont {V.~K.}\ \bibnamefont {Dugaev}}, \bibinfo {author} {\bibfnamefont {C.-L.}\ \bibnamefont {Jia}}, \bibinfo {author} {\bibfnamefont {A.}~\bibnamefont {Ernst}}, \bibinfo {author} {\bibfnamefont {A.}~\bibnamefont {Komnik}},\ and\ \bibinfo {author} {\bibfnamefont {J.}~\bibnamefont {Berakdar}},\ }\bibfield  {title} {\bibinfo {title} {Topological insulator in a helicoidal magnetization field},\ }\href {https://doi.org/10.1103/PhysRevB.94.174436} {\bibfield  {journal} {\bibinfo  {journal} {Phys. Rev. B}\ }\textbf {\bibinfo {volume} {94}},\ \bibinfo {pages} {174436} (\bibinfo {year} {2016})}\BibitemShut {NoStop}%
\bibitem [{\citenamefont {Li}\ \emph {et~al.}(2011)\citenamefont {Li}, \citenamefont {Ghosh}, \citenamefont {Sau}, \citenamefont {Tewari},\ and\ \citenamefont {Das~Sarma}}]{Li2011}%
  \BibitemOpen
  \bibfield  {author} {\bibinfo {author} {\bibfnamefont {Q.}~\bibnamefont {Li}}, \bibinfo {author} {\bibfnamefont {P.}~\bibnamefont {Ghosh}}, \bibinfo {author} {\bibfnamefont {J.~D.}\ \bibnamefont {Sau}}, \bibinfo {author} {\bibfnamefont {S.}~\bibnamefont {Tewari}},\ and\ \bibinfo {author} {\bibfnamefont {S.}~\bibnamefont {Das~Sarma}},\ }\bibfield  {title} {\bibinfo {title} {Anisotropic surface transport in topological insulators in proximity to a helical spin density wave},\ }\href {https://doi.org/10.1103/PhysRevB.83.085110} {\bibfield  {journal} {\bibinfo  {journal} {Phys. Rev. B}\ }\textbf {\bibinfo {volume} {83}},\ \bibinfo {pages} {085110} (\bibinfo {year} {2011})}\BibitemShut {NoStop}%
\bibitem [{\citenamefont {Wakabayashi}\ \emph {et~al.}(2009)\citenamefont {Wakabayashi}, \citenamefont {Takane}, \citenamefont {Yamamoto},\ and\ \citenamefont {Sigrist}}]{Wakabayashi2009}%
  \BibitemOpen
  \bibfield  {author} {\bibinfo {author} {\bibfnamefont {K.}~\bibnamefont {Wakabayashi}}, \bibinfo {author} {\bibfnamefont {Y.}~\bibnamefont {Takane}}, \bibinfo {author} {\bibfnamefont {M.}~\bibnamefont {Yamamoto}},\ and\ \bibinfo {author} {\bibfnamefont {M.}~\bibnamefont {Sigrist}},\ }\bibfield  {title} {\bibinfo {title} {Electronic transport properties of graphene nanoribbons},\ }\href {https://doi.org/10.1088/1367-2630/11/9/095016} {\bibfield  {journal} {\bibinfo  {journal} {New J. Phys.}\ }\textbf {\bibinfo {volume} {11}},\ \bibinfo {pages} {095016} (\bibinfo {year} {2009})}\BibitemShut {NoStop}%
\bibitem [{\citenamefont {Manfrinato}\ \emph {et~al.}(2013)\citenamefont {Manfrinato}, \citenamefont {Zhang}, \citenamefont {Su}, \citenamefont {Duan}, \citenamefont {Hobbs}, \citenamefont {Stach},\ and\ \citenamefont {Berggren}}]{EBL2013}%
  \BibitemOpen
  \bibfield  {author} {\bibinfo {author} {\bibfnamefont {V.~R.}\ \bibnamefont {Manfrinato}}, \bibinfo {author} {\bibfnamefont {L.}~\bibnamefont {Zhang}}, \bibinfo {author} {\bibfnamefont {D.}~\bibnamefont {Su}}, \bibinfo {author} {\bibfnamefont {H.}~\bibnamefont {Duan}}, \bibinfo {author} {\bibfnamefont {R.~G.}\ \bibnamefont {Hobbs}}, \bibinfo {author} {\bibfnamefont {E.~A.}\ \bibnamefont {Stach}},\ and\ \bibinfo {author} {\bibfnamefont {K.~K.}\ \bibnamefont {Berggren}},\ }\bibfield  {title} {\bibinfo {title} {Resolution limits of electron-beam lithography toward the atomic scale},\ }\href {https://doi.org/10.1021/nl304715p} {\bibfield  {journal} {\bibinfo  {journal} {Nano Lett.}\ }\textbf {\bibinfo {volume} {13}},\ \bibinfo {pages} {1555} (\bibinfo {year} {2013})}\BibitemShut {NoStop}%
\bibitem [{\citenamefont {Gudmundsson}\ \emph {et~al.}(2022)\citenamefont {Gudmundsson}, \citenamefont {Anders},\ and\ \citenamefont {von Keudell}}]{PVD_2022}%
  \BibitemOpen
  \bibfield  {author} {\bibinfo {author} {\bibfnamefont {J.~T.}\ \bibnamefont {Gudmundsson}}, \bibinfo {author} {\bibfnamefont {A.}~\bibnamefont {Anders}},\ and\ \bibinfo {author} {\bibfnamefont {A.}~\bibnamefont {von Keudell}},\ }\bibfield  {title} {\bibinfo {title} {Foundations of physical vapor deposition with plasma assistance},\ }\href {https://doi.org/10.1088/1361-6595/ac7f53} {\bibfield  {journal} {\bibinfo  {journal} {Plasma Sources Sci. Technol.}\ }\textbf {\bibinfo {volume} {31}},\ \bibinfo {pages} {083001} (\bibinfo {year} {2022})}\BibitemShut {NoStop}%
\bibitem [{Note1()}]{Note1}%
  \BibitemOpen
  \bibinfo {note} {To incorporate edge effects, we take hopping to edge sites $t'=0.9t$, with $t$ hopping in bulk sites}\BibitemShut {NoStop}%
\bibitem [{\citenamefont {Kane}\ and\ \citenamefont {Mele}(2005)}]{KaneMele2005}%
  \BibitemOpen
  \bibfield  {author} {\bibinfo {author} {\bibfnamefont {C.~L.}\ \bibnamefont {Kane}}\ and\ \bibinfo {author} {\bibfnamefont {E.~J.}\ \bibnamefont {Mele}},\ }\bibfield  {title} {\bibinfo {title} {Quantum spin {H}all effect in graphene},\ }\href {https://doi.org/10.1103/PhysRevLett.95.226801} {\bibfield  {journal} {\bibinfo  {journal} {Phys. Rev. Lett.}\ }\textbf {\bibinfo {volume} {95}},\ \bibinfo {pages} {226801} (\bibinfo {year} {2005})}\BibitemShut {NoStop}%
\bibitem [{\citenamefont {Christensen}\ \emph {et~al.}(2016)\citenamefont {Christensen}, \citenamefont {Schecter}, \citenamefont {Flensberg}, \citenamefont {Andersen},\ and\ \citenamefont {Paaske}}]{PhysRevB.94.144509}%
  \BibitemOpen
  \bibfield  {author} {\bibinfo {author} {\bibfnamefont {M.~H.}\ \bibnamefont {Christensen}}, \bibinfo {author} {\bibfnamefont {M.}~\bibnamefont {Schecter}}, \bibinfo {author} {\bibfnamefont {K.}~\bibnamefont {Flensberg}}, \bibinfo {author} {\bibfnamefont {B.~M.}\ \bibnamefont {Andersen}},\ and\ \bibinfo {author} {\bibfnamefont {J.}~\bibnamefont {Paaske}},\ }\bibfield  {title} {\bibinfo {title} {Spiral magnetic order and topological superconductivity in a chain of magnetic adatoms on a two-dimensional superconductor},\ }\href {https://doi.org/10.1103/PhysRevB.94.144509} {\bibfield  {journal} {\bibinfo  {journal} {Phys. Rev. B}\ }\textbf {\bibinfo {volume} {94}},\ \bibinfo {pages} {144509} (\bibinfo {year} {2016})}\BibitemShut {NoStop}%
\bibitem [{\citenamefont {Chen}\ \emph {et~al.}(2022{\natexlab{b}})\citenamefont {Chen}, \citenamefont {Khosravian}, \citenamefont {Lado},\ and\ \citenamefont {Ramires}}]{Chen2022spiral}%
  \BibitemOpen
  \bibfield  {author} {\bibinfo {author} {\bibfnamefont {G.}~\bibnamefont {Chen}}, \bibinfo {author} {\bibfnamefont {M.}~\bibnamefont {Khosravian}}, \bibinfo {author} {\bibfnamefont {J.~L.}\ \bibnamefont {Lado}},\ and\ \bibinfo {author} {\bibfnamefont {A.}~\bibnamefont {Ramires}},\ }\bibfield  {title} {\bibinfo {title} {Designing spin-textured flat bands in twisted graphene multilayers via helimagnet encapsulation},\ }\href {https://doi.org/10.1088/2053-1583/ac4af8} {\bibfield  {journal} {\bibinfo  {journal} {2D Materials}\ }\textbf {\bibinfo {volume} {9}},\ \bibinfo {pages} {024002} (\bibinfo {year} {2022}{\natexlab{b}})}\BibitemShut {NoStop}%
\bibitem [{\citenamefont {Brekke}\ \emph {et~al.}(2024)\citenamefont {Brekke}, \citenamefont {Sukhachov}, \citenamefont {Giil}, \citenamefont {Brataas},\ and\ \citenamefont {Linder}}]{PhysRevLett.133.236703}%
  \BibitemOpen
  \bibfield  {author} {\bibinfo {author} {\bibfnamefont {B.}~\bibnamefont {Brekke}}, \bibinfo {author} {\bibfnamefont {P.}~\bibnamefont {Sukhachov}}, \bibinfo {author} {\bibfnamefont {H.~G.}\ \bibnamefont {Giil}}, \bibinfo {author} {\bibfnamefont {A.}~\bibnamefont {Brataas}},\ and\ \bibinfo {author} {\bibfnamefont {J.}~\bibnamefont {Linder}},\ }\bibfield  {title} {\bibinfo {title} {Minimal models and transport properties of unconventional $p$-wave magnets},\ }\href {https://doi.org/10.1103/PhysRevLett.133.236703} {\bibfield  {journal} {\bibinfo  {journal} {Phys. Rev. Lett.}\ }\textbf {\bibinfo {volume} {133}},\ \bibinfo {pages} {236703} (\bibinfo {year} {2024})}\BibitemShut {NoStop}%
\bibitem [{\citenamefont {Song}\ \emph {et~al.}(2021)\citenamefont {Song}, \citenamefont {Kang}, \citenamefont {Rhodes}, \citenamefont {Kim}, \citenamefont {Hone},\ and\ \citenamefont {Ryu}}]{Song_2021}%
  \BibitemOpen
  \bibfield  {author} {\bibinfo {author} {\bibfnamefont {M.}~\bibnamefont {Song}}, \bibinfo {author} {\bibfnamefont {H.}~\bibnamefont {Kang}}, \bibinfo {author} {\bibfnamefont {D.}~\bibnamefont {Rhodes}}, \bibinfo {author} {\bibfnamefont {B.}~\bibnamefont {Kim}}, \bibinfo {author} {\bibfnamefont {J.}~\bibnamefont {Hone}},\ and\ \bibinfo {author} {\bibfnamefont {S.}~\bibnamefont {Ryu}},\ }\bibfield  {title} {\bibinfo {title} {Optically facet-resolved reaction anisotropy in two-dimensional transition metal dichalcogenides},\ }\href {https://doi.org/10.1088/2053-1583/ac0297} {\bibfield  {journal} {\bibinfo  {journal} {2D Mater.}\ }\textbf {\bibinfo {volume} {8}},\ \bibinfo {pages} {035045} (\bibinfo {year} {2021})}\BibitemShut {NoStop}%
\bibitem [{\citenamefont {Wang}\ \emph {et~al.}(2024{\natexlab{b}})\citenamefont {Wang}, \citenamefont {Chen}, \citenamefont {Lin}, \citenamefont {Lin}, \citenamefont {Chang}, \citenamefont {Muthu}, \citenamefont {Hofmann},\ and\ \citenamefont {Hsieh}}]{Wang_2024}%
  \BibitemOpen
  \bibfield  {author} {\bibinfo {author} {\bibfnamefont {H.}~\bibnamefont {Wang}}, \bibinfo {author} {\bibfnamefont {D.-R.}\ \bibnamefont {Chen}}, \bibinfo {author} {\bibfnamefont {Y.-C.}\ \bibnamefont {Lin}}, \bibinfo {author} {\bibfnamefont {P.-H.}\ \bibnamefont {Lin}}, \bibinfo {author} {\bibfnamefont {J.-T.}\ \bibnamefont {Chang}}, \bibinfo {author} {\bibfnamefont {J.}~\bibnamefont {Muthu}}, \bibinfo {author} {\bibfnamefont {M.}~\bibnamefont {Hofmann}},\ and\ \bibinfo {author} {\bibfnamefont {Y.-P.}\ \bibnamefont {Hsieh}},\ }\bibfield  {title} {\bibinfo {title} {Enhancing the electrochemical activity of 2{D} materials edges through oriented electric fields},\ }\href {https://doi.org/10.1021/acsnano.4c06341} {\bibfield  {journal} {\bibinfo  {journal} {ACS Nano}\ }\textbf {\bibinfo {volume} {18}},\ \bibinfo {pages} {19828} (\bibinfo {year} {2024}{\natexlab{b}})}\BibitemShut {NoStop}%
\bibitem [{\citenamefont {Wu}\ \emph {et~al.}(2016)\citenamefont {Wu}, \citenamefont {Song}, \citenamefont {Zhou},\ and\ \citenamefont {Jiang}}]{wu2016disorder}%
  \BibitemOpen
  \bibfield  {author} {\bibinfo {author} {\bibfnamefont {B.}~\bibnamefont {Wu}}, \bibinfo {author} {\bibfnamefont {J.}~\bibnamefont {Song}}, \bibinfo {author} {\bibfnamefont {J.}~\bibnamefont {Zhou}},\ and\ \bibinfo {author} {\bibfnamefont {H.}~\bibnamefont {Jiang}},\ }\bibfield  {title} {\bibinfo {title} {Disorder effects in topological states: Brief review of the recent developments},\ }\href {https://doi.org/10.1088/1674-1056/25/11/117311} {\bibfield  {journal} {\bibinfo  {journal} {Chin. Phys. B}\ }\textbf {\bibinfo {volume} {25}},\ \bibinfo {pages} {117311} (\bibinfo {year} {2016})}\BibitemShut {NoStop}%
\bibitem [{\citenamefont {Gupta}\ and\ \citenamefont {Srivastava}(2020)}]{Gupta2020}%
  \BibitemOpen
  \bibfield  {author} {\bibinfo {author} {\bibfnamefont {A.}~\bibnamefont {Gupta}}\ and\ \bibinfo {author} {\bibfnamefont {S.~K.}\ \bibnamefont {Srivastava}},\ }\bibfield  {title} {\bibinfo {title} {Paramagnetism, hopping conduction, and weak localization in highly disordered pure and {D}y-doped {B}i2{S}e3 nanoplates},\ }\href {https://doi.org/10.1063/1.5140412} {\bibfield  {journal} {\bibinfo  {journal} {J. Appl. Phys.}\ }\textbf {\bibinfo {volume} {127}},\ \bibinfo {pages} {244302} (\bibinfo {year} {2020})}\BibitemShut {NoStop}%
\bibitem [{\citenamefont {Wang}\ \emph {et~al.}(2019)\citenamefont {Wang}, \citenamefont {Zhou}, \citenamefont {Jiang}, \citenamefont {Sun}, \citenamefont {Zang}, \citenamefont {Gong}, \citenamefont {Zhang}, \citenamefont {Tong}, \citenamefont {Xie}, \citenamefont {Liu}, \citenamefont {Chen}, \citenamefont {He},\ and\ \citenamefont {Xue}}]{Wang2019}%
  \BibitemOpen
  \bibfield  {author} {\bibinfo {author} {\bibfnamefont {Z.}~\bibnamefont {Wang}}, \bibinfo {author} {\bibfnamefont {T.}~\bibnamefont {Zhou}}, \bibinfo {author} {\bibfnamefont {T.}~\bibnamefont {Jiang}}, \bibinfo {author} {\bibfnamefont {H.}~\bibnamefont {Sun}}, \bibinfo {author} {\bibfnamefont {Y.}~\bibnamefont {Zang}}, \bibinfo {author} {\bibfnamefont {Y.}~\bibnamefont {Gong}}, \bibinfo {author} {\bibfnamefont {J.}~\bibnamefont {Zhang}}, \bibinfo {author} {\bibfnamefont {M.}~\bibnamefont {Tong}}, \bibinfo {author} {\bibfnamefont {X.}~\bibnamefont {Xie}}, \bibinfo {author} {\bibfnamefont {Q.}~\bibnamefont {Liu}}, \bibinfo {author} {\bibfnamefont {C.}~\bibnamefont {Chen}}, \bibinfo {author} {\bibfnamefont {K.}~\bibnamefont {He}},\ and\ \bibinfo {author} {\bibfnamefont {Q.-K.}\ \bibnamefont {Xue}},\ }\bibfield  {title} {\bibinfo {title} {Dimensional crossover and topological nature of the thin films of a three-dimensional topological insulator by band gap engineering},\ }\href
  {https://doi.org/10.1021/acs.nanolett.9b01641} {\bibfield  {journal} {\bibinfo  {journal} {Nano Lett.}\ }\textbf {\bibinfo {volume} {19}},\ \bibinfo {pages} {4627} (\bibinfo {year} {2019})}\BibitemShut {NoStop}%
\bibitem [{\citenamefont {Datta}(1997)}]{Datta1997}%
  \BibitemOpen
  \bibfield  {author} {\bibinfo {author} {\bibfnamefont {S.}~\bibnamefont {Datta}},\ }\href@noop {} {\emph {\bibinfo {title} {Electronic transport in mesoscopic systems}}}\ (\bibinfo  {publisher} {Cambridge university press},\ \bibinfo {address} {Cambridge},\ \bibinfo {year} {1997})\BibitemShut {NoStop}%
\bibitem [{pyq()}]{pyqula}%
  \BibitemOpen
  \href@noop {} {\bibinfo {title} {P{Y}{Q}{U}{L}{A} {L}ibrary}},\ \bibinfo {howpublished} {\url{https://github.com/joselado/pyqula}}\BibitemShut {NoStop}%
\bibitem [{\citenamefont {Shafiei}\ \emph {et~al.}(2022)\citenamefont {Shafiei}, \citenamefont {Fazileh}, \citenamefont {Peeters},\ and\ \citenamefont {Milo\ifmmode \check{s}\else \v{s}\fi{}evi\ifmmode~\acute{c}\else \'{c}\fi{}}}]{Shafiei2022}%
  \BibitemOpen
  \bibfield  {author} {\bibinfo {author} {\bibfnamefont {M.}~\bibnamefont {Shafiei}}, \bibinfo {author} {\bibfnamefont {F.}~\bibnamefont {Fazileh}}, \bibinfo {author} {\bibfnamefont {F.~M.}\ \bibnamefont {Peeters}},\ and\ \bibinfo {author} {\bibfnamefont {M.~V.}\ \bibnamefont {Milo\ifmmode \check{s}\else \v{s}\fi{}evi\ifmmode~\acute{c}\else \'{c}\fi{}}},\ }\bibfield  {title} {\bibinfo {title} {Controlling the hybridization gap and transport in a thin-film topological insulator: Effect of strain, and electric and magnetic field},\ }\href {https://doi.org/10.1103/PhysRevB.106.035119} {\bibfield  {journal} {\bibinfo  {journal} {Phys. Rev. B}\ }\textbf {\bibinfo {volume} {106}},\ \bibinfo {pages} {035119} (\bibinfo {year} {2022})}\BibitemShut {NoStop}%
\bibitem [{\citenamefont {Fatemi}\ \emph {et~al.}(2018)\citenamefont {Fatemi}, \citenamefont {Wu}, \citenamefont {Cao}, \citenamefont {Bretheau}, \citenamefont {Gibson}, \citenamefont {Watanabe}, \citenamefont {Taniguchi}, \citenamefont {Cava},\ and\ \citenamefont {Jarillo-Herrero}}]{Fatemi2018}%
  \BibitemOpen
  \bibfield  {author} {\bibinfo {author} {\bibfnamefont {V.}~\bibnamefont {Fatemi}}, \bibinfo {author} {\bibfnamefont {S.}~\bibnamefont {Wu}}, \bibinfo {author} {\bibfnamefont {Y.}~\bibnamefont {Cao}}, \bibinfo {author} {\bibfnamefont {L.}~\bibnamefont {Bretheau}}, \bibinfo {author} {\bibfnamefont {Q.~D.}\ \bibnamefont {Gibson}}, \bibinfo {author} {\bibfnamefont {K.}~\bibnamefont {Watanabe}}, \bibinfo {author} {\bibfnamefont {T.}~\bibnamefont {Taniguchi}}, \bibinfo {author} {\bibfnamefont {R.~J.}\ \bibnamefont {Cava}},\ and\ \bibinfo {author} {\bibfnamefont {P.}~\bibnamefont {Jarillo-Herrero}},\ }\bibfield  {title} {\bibinfo {title} {Electrically tunable low-density superconductivity in a monolayer topological insulator},\ }\href {https://doi.org/10.1126/science.aar4642} {\bibfield  {journal} {\bibinfo  {journal} {Science}\ }\textbf {\bibinfo {volume} {362}},\ \bibinfo {pages} {926–929} (\bibinfo {year} {2018})}\BibitemShut {NoStop}%
\bibitem [{\citenamefont {Abrahams}\ \emph {et~al.}(1979)\citenamefont {Abrahams}, \citenamefont {Anderson}, \citenamefont {Licciardello},\ and\ \citenamefont {Ramakrishnan}}]{Abrahams1979}%
  \BibitemOpen
  \bibfield  {author} {\bibinfo {author} {\bibfnamefont {E.}~\bibnamefont {Abrahams}}, \bibinfo {author} {\bibfnamefont {P.~W.}\ \bibnamefont {Anderson}}, \bibinfo {author} {\bibfnamefont {D.~C.}\ \bibnamefont {Licciardello}},\ and\ \bibinfo {author} {\bibfnamefont {T.~V.}\ \bibnamefont {Ramakrishnan}},\ }\bibfield  {title} {\bibinfo {title} {Scaling theory of localization: Absence of quantum diffusion in two dimensions},\ }\href {https://doi.org/10.1103/PhysRevLett.42.673} {\bibfield  {journal} {\bibinfo  {journal} {Phys. Rev. Lett.}\ }\textbf {\bibinfo {volume} {42}},\ \bibinfo {pages} {673} (\bibinfo {year} {1979})}\BibitemShut {NoStop}%
\bibitem [{\citenamefont {Yang}\ \emph {et~al.}(2020)\citenamefont {Yang}, \citenamefont {Li}, \citenamefont {Yin}, \citenamefont {Xu}, \citenamefont {Mullan}, \citenamefont {Taniguchi}, \citenamefont {Watanabe}, \citenamefont {Geim}, \citenamefont {Novoselov},\ and\ \citenamefont {Mishchenko}}]{Yang2020}%
  \BibitemOpen
  \bibfield  {author} {\bibinfo {author} {\bibfnamefont {Y.}~\bibnamefont {Yang}}, \bibinfo {author} {\bibfnamefont {J.}~\bibnamefont {Li}}, \bibinfo {author} {\bibfnamefont {J.}~\bibnamefont {Yin}}, \bibinfo {author} {\bibfnamefont {S.}~\bibnamefont {Xu}}, \bibinfo {author} {\bibfnamefont {C.}~\bibnamefont {Mullan}}, \bibinfo {author} {\bibfnamefont {T.}~\bibnamefont {Taniguchi}}, \bibinfo {author} {\bibfnamefont {K.}~\bibnamefont {Watanabe}}, \bibinfo {author} {\bibfnamefont {A.~K.}\ \bibnamefont {Geim}}, \bibinfo {author} {\bibfnamefont {K.~S.}\ \bibnamefont {Novoselov}},\ and\ \bibinfo {author} {\bibfnamefont {A.}~\bibnamefont {Mishchenko}},\ }\bibfield  {title} {\bibinfo {title} {In situ manipulation of van der {W}aals heterostructures for twistronics},\ }\href {http://dx.doi.org/10.1126/sciadv.abd3655} {\bibfield  {journal} {\bibinfo  {journal} {Sci. Adv.}\ }\textbf {\bibinfo {volume} {6}},\ \bibinfo {pages} {eabd3655} (\bibinfo {year} {2020})}\BibitemShut {NoStop}%
\bibitem [{\citenamefont {Wirth}\ and\ \citenamefont {Steglich}(2016)}]{Wirth2016}%
  \BibitemOpen
  \bibfield  {author} {\bibinfo {author} {\bibfnamefont {S.}~\bibnamefont {Wirth}}\ and\ \bibinfo {author} {\bibfnamefont {F.}~\bibnamefont {Steglich}},\ }\bibfield  {title} {\bibinfo {title} {Exploring heavy fermions from macroscopic to microscopic length scales},\ }\href {http://dx.doi.org/10.1038/natrevmats.2016.51} {\bibfield  {journal} {\bibinfo  {journal} {Nat. Rev. Mater.}\ }\textbf {\bibinfo {volume} {1}},\ \bibinfo {pages} {16051} (\bibinfo {year} {2016})}\BibitemShut {NoStop}%
\bibitem [{\citenamefont {{Lippo}}\ \emph {et~al.}()\citenamefont {{Lippo}}, \citenamefont {{Pereira}}, \citenamefont {{Lado}},\ and\ \citenamefont {{Chen}}}]{2024arXiv240917202L}%
  \BibitemOpen
  \bibfield  {author} {\bibinfo {author} {\bibfnamefont {Z.}~\bibnamefont {{Lippo}}}, \bibinfo {author} {\bibfnamefont {E.~L.}\ \bibnamefont {{Pereira}}}, \bibinfo {author} {\bibfnamefont {J.~L.}\ \bibnamefont {{Lado}}},\ and\ \bibinfo {author} {\bibfnamefont {G.}~\bibnamefont {{Chen}}},\ }\bibfield  {title} {\bibinfo {title} {{Topological zero modes and correlation pumping in an engineered {K}ondo lattice}},\ }\href@noop {} {\ }\Eprint {https://arxiv.org/abs/2409.17202} {arXiv:2409.17202} \BibitemShut {NoStop}%
\bibitem [{\citenamefont {Chen}\ \emph {et~al.}(2024)\citenamefont {Chen}, \citenamefont {Stoudenmire}, \citenamefont {Komijani},\ and\ \citenamefont {Coleman}}]{PhysRevResearch.6.023227}%
  \BibitemOpen
  \bibfield  {author} {\bibinfo {author} {\bibfnamefont {J.}~\bibnamefont {Chen}}, \bibinfo {author} {\bibfnamefont {E.~M.}\ \bibnamefont {Stoudenmire}}, \bibinfo {author} {\bibfnamefont {Y.}~\bibnamefont {Komijani}},\ and\ \bibinfo {author} {\bibfnamefont {P.}~\bibnamefont {Coleman}},\ }\bibfield  {title} {\bibinfo {title} {Matrix product study of spin fractionalization in the one-dimensional {K}ondo insulator},\ }\href {https://doi.org/10.1103/PhysRevResearch.6.023227} {\bibfield  {journal} {\bibinfo  {journal} {Phys. Rev. Res.}\ }\textbf {\bibinfo {volume} {6}},\ \bibinfo {pages} {023227} (\bibinfo {year} {2024})}\BibitemShut {NoStop}%
\bibitem [{\citenamefont {Lee}\ \emph {et~al.}(2018)\citenamefont {Lee}, \citenamefont {Richardella}, \citenamefont {Fraleigh}, \citenamefont {Liu}, \citenamefont {Zhao},\ and\ \citenamefont {Samarth}}]{Lee2018}%
  \BibitemOpen
  \bibfield  {author} {\bibinfo {author} {\bibfnamefont {J.~S.}\ \bibnamefont {Lee}}, \bibinfo {author} {\bibfnamefont {A.}~\bibnamefont {Richardella}}, \bibinfo {author} {\bibfnamefont {R.~D.}\ \bibnamefont {Fraleigh}}, \bibinfo {author} {\bibfnamefont {C.-x.}\ \bibnamefont {Liu}}, \bibinfo {author} {\bibfnamefont {W.}~\bibnamefont {Zhao}},\ and\ \bibinfo {author} {\bibfnamefont {N.}~\bibnamefont {Samarth}},\ }\bibfield  {title} {\bibinfo {title} {Engineering the breaking of time-reversal symmetry in gate-tunable hybrid ferromagnet/topological insulator heterostructures},\ }\href {https://doi.org/10.1038/s41535-018-0123-2} {\bibfield  {journal} {\bibinfo  {journal} {npj Quantum Mater.}\ }\textbf {\bibinfo {volume} {3}},\ \bibinfo {pages} {51} (\bibinfo {year} {2018})}\BibitemShut {NoStop}%
\bibitem [{\citenamefont {Nigmatulin}\ \emph {et~al.}()\citenamefont {Nigmatulin}, \citenamefont {Lado},\ and\ \citenamefont {Sun}}]{nigmatulin_2025_15755975}%
  \BibitemOpen
  \bibfield  {author} {\bibinfo {author} {\bibfnamefont {F.}~\bibnamefont {Nigmatulin}}, \bibinfo {author} {\bibfnamefont {J.~L.}\ \bibnamefont {Lado}},\ and\ \bibinfo {author} {\bibfnamefont {Z.}~\bibnamefont {Sun}},\ }\bibfield  {title} {\bibinfo {title} {Electrical probe of spin-spiral order in quantum spin {H}all spin-spiral magnet van der {W}aals heterostructures ({D}ata)},\ }\href {https://zenodo.org/records/15755975} {https://zenodo.org/records/15755975},\ \bibinfo {note} {{V}ersion v1, Zenodo (2025)}\BibitemShut {NoStop}%
\end{thebibliography}%

\end{document}